\documentclass[
preprint,
 amsmath,amssymb,
 aps,
]{revtex4-2}

\usepackage{graphicx}
\usepackage{dcolumn}
\usepackage{bm}
\usepackage[colorlinks=true,linkcolor=blue,citecolor=blue,urlcolor=blue]{hyperref}
\usepackage[nameinlink]{cleveref}
\usepackage{subfigure}

\renewcommand{\d}[2]{\frac{\text{d} #1}{\text{d} #2}}

\begin{document}

\preprint{APS/123-QED}

\title{Group cohesion and passive dynamics of a pair of inertial swimmers with three-dimensional hydrodynamic interactions}

\author{Mohamed Niged Mabrouk}
\email{mmabrouk@uh.edu}
\author{Daniel Floryan}%
\email{dfloryan@uh.edu}
\affiliation{%
 Department of Mechanical Engineering, University of Houston, Houston, TX 77204, USA
}%

\date{\today}

\begin{abstract}
When swimming animals form cohesive groups, they can reap several benefits. Our understanding of collective animal motion has traditionally been driven by models based on phenomenological behavioral rules, but more recent work has highlighted the critical importance of hydrodynamic interactions among a group of inertial swimmers. To study how hydrodynamic interactions affect group cohesion, we develop a three-dimensional, inviscid, far-field model of a swimmer. In a group of two model swimmers, we observe several dynamical phases, including following, divergence, collision, and cohesion. Our results illustrate when cohesive groups can passively form through hydrodynamic interactions alone, and when other action is needed to maintain cohesion. We find that misalignment between swimmers makes passive cohesion less likely; nevertheless, it is possible for a cohesive group to form through passive hydrodynamic interactions alone. We also find that the geometry of swimmers critically affects the group dynamics due to its role in how swimmers sample the velocity gradient of the flow. 
\end{abstract}

\maketitle


\section{\label{sec1:intro}Introduction}
The collective motion of animals can be seen in numerous examples throughout the animal kingdom \cite{vicsek2012collective}. A canonical example is a large fish school, which can be defined as a structured and coordinated group of swimming organisms with synchronized movements \cite{e3c99d1b-f54c-39bd-90da-8cd2a5ff32ba}. Such collective motion of swimming animals confers several benefits, including defence against predators \cite{major1978predator, shaw1978schooling, landeau1986oddity}, enhanced foraging success \cite{pitcher1982fish}, mating success \cite{barnes1999introduction}, and increased swimming efficiency \cite{weihs1973hydromechanics}. Furthermore, groups are robust in the sense that if one individual leaves the group, the group can still retain its advantages \cite{brambilla2013swarm}. The benefits of collective motion have even led roboticists to develop robot swarms that aim to attain similar functionality \cite{rubenstein2014programmable, dorigo2020reflections}, but maintaining a cohesive group is a difficult task that requires unique strategies to coordinate dynamically interacting individuals \cite{gordon2010ant}. We are principally interested in this aspect of schooling---the cohesion of groups---especially in the role that hydrodynamic interactions play. 

Since \citeauthor{weihs1973hydromechanics}' seminal work more than 50 years ago \cite{weihs1973hydromechanics}, substantial work has been done to understand how inertial swimmers may harness hydrodynamic interactions between each other when schooling \cite{li2020vortex, verma2018efficient, ashraf2017simple, novati2017synchronisation, oza2019lattices, alben2021collective1, alben2021collective2, newbolt2022lateral, kurt2018flow, boschitsch2014propulsive, dewey2014propulsive, pan2020computational, muscutt2017performance, seo2022improved, pan2022effects, saadat2021hydrodynamic}. Synthesizing these studies, schooling can lead to reduced energy expenditure when swimming under the right conditions. A swimmer can reduce its energy expenditure by maintaining an appropriate phase offset between its kinematics and those of its neighbors, with the phase offset depending on the relative location of a swimmer to its neighbor. The phase relations arise as a result of direct interactions between two swimmers, or due to the interaction between a follower and the vortical wake of the leader. In dense schools, an effect similar to ground effect may arise \cite{pan2020computational}. In addition to energy savings, some of the cited works also note that schooling can increase swimming speed relative to swimming in isolation. 

The hydrodynamic (and other) benefits of schooling can only be realized if the appropriate group structure is maintained. Thus, the dynamics that dictate the motions of freely moving swimmers become important. These dynamics are governed by the active decision-making and social interactions of swimmers \cite{couzin2003self, couzin2005effective}, as well as by passive hydrodynamic interactions. A substantial and important body of work on active decision-making and social interactions exists, and it has produced a variety of agent-based models for the interactions between individuals in a group \cite{breder1954equations, 19821081, reynolds1987flocks, couzin2002collective}. These models range in complexity, but they tend to neglect the role of hydrodynamic interactions, which are expected to have an outsized effect on the collective motion of swimmers \cite{ko2023role}. Here, we focus on passive hydrodynamic interactions. Considering only the hydrodynamic interactions allows us to answer whether a cohesive group can form passively, or whether feedback control is required to maintain a cohesive group. Additionally, a deeper understanding of the hydrodynamic interactions may change the inferences that one makes about social interactions between animals based on observations of groups of animals, leading to improved agent-based models \cite{filella2018model}, and could also lead to new physics-based models of underwater robots that could be used for improved model-based control of groups. 

\citeauthor{lighthill1975mathematical} famously raised the possibility that the order observed within groups in nature may passively arise due to flow-mediated interactions between group members \cite{lighthill1975mathematical}. \citeauthor{kelly1959two}'s earlier analysis of two spheres moving with equal and constant velocities in potential flow lent some credence to this possibility as it showed that the spheres can experience attractive forces, depending on their configuration \cite{kelly1959two}. Several studies have sought to understand the fluid-mediated dynamics of multiple swimmers interacting with each other \cite{tchieu2012finite, kanso2014dipole, gazzola2016learning, filella2018model, porfiri2022hydrodynamic, becker2015hydrodynamic, ramananarivo2016flow, newbolt2019flow, oza2019lattices, newbolt2022lateral, gazzola2014reinforcement}, with nearly all of them being two-dimensional in nature. Ristroph's work has consistently shown that swimmers in schools passively and stably arrange themselves so that their motion has a particular phase offset from incoming vortical wakes \cite{becker2015hydrodynamic, ramananarivo2016flow, newbolt2019flow}. It is important to note that the experiments leading to this conclusion were highly constrained, with each model swimmer having only one unconstrained degree of freedom. Additional unconstrained degrees of freedom may qualitatively change the dynamics, as shown in \cite{lin2021flow, lin2022two, ormonde2021two}, where the two translational degrees of freedom were unconstrained. We expect that unconstrained rotational motion will lead to further changes in the dynamics. 

Simplified vortex-dipole models that are still two-dimensional but do not constrain the degrees of freedom have been developed to study the cohesion of groups of swimmers \cite{tchieu2012finite, kanso2014dipole, gazzola2016learning, filella2018model}. According to these models, purely hydrodynamic interactions are not always enough to maintain cohesive groups; active control on the part of the swimmers is required.

It is unknown how three-dimensional hydrodynamic interactions, as well as the additional degrees of freedom of a three-dimensional world, affect the stability and cohesion of schools. However, it is immediately clear that the three-dimensional dynamics will be different from the two-dimensional dynamics since the velocity field of a three-dimensional swimmer decays with distance $r$ as $r^{-3}$, while the velocity field of a two-dimensional swimmer decays as $r^{-2}$. Any potential differences are important since our current
understanding of the role of hydrodynamics in collective motion is based almost solely on the two-dimensional scenario.

In this work, we take the first step towards understanding the passive group dynamics that develop under three-dimensional hydrodynamic interactions between swimmers. This understanding is enabled by the development of a simple model of a freely moving inertial swimmer that captures the correct leading-order hydrodynamic interactions between swimmers. We consider the elemental interactions between a pair of swimmers. By focusing on their planar dynamics, we are able to conduct a comparative analysis between two-dimensional and three-dimensional swimmers, thereby revealing the significance of three-dimensionality in schooling behavior. Symmetry reduction allows us to fully characterize the possible planar dynamics. Additionally, we demonstrate that the non-planar dynamics of a pair are of the same character as the planar dynamics, and that it is possible for a group to be passively cohesive in three dimensions. Finally, we make observations on how the geometry of a swimmer affects the dynamics by comparing swimmers that are long and thin to those that are short and fat.

\section{\label{sec2:math}Mathematical Model} 
At the large Reynolds numbers possessed by inertial swimmers, the flow away from their bodies is well described by potential flow. Here, we develop a potential flow model of hydrodynamic interactions between swimmers. First, we describe the flow induced by the motion of one swimmer. Then, we develop a model for the dynamics of multiple swimmers.

\subsection{Flow induced by one swimmer}

\begin{figure}
  \centering
  \subfigure[]{
  \includegraphics[width=0.3\textwidth]{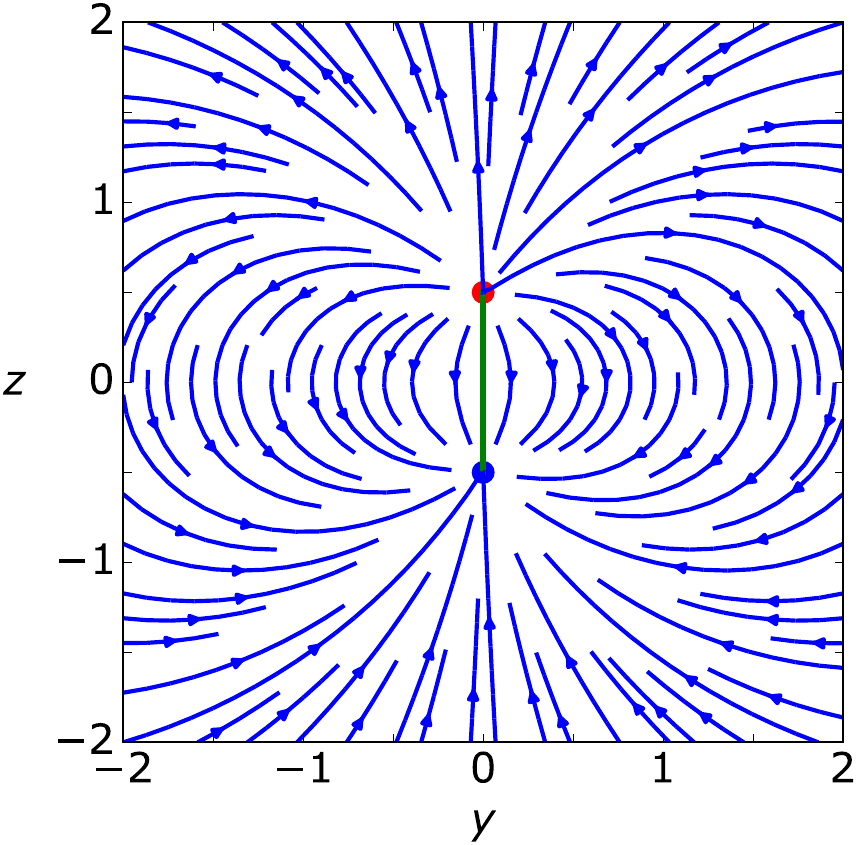}
  \label{fig1a:StreamlinesPlanar}
  }
  \subfigure[]{
    \includegraphics[width=0.3\textwidth]{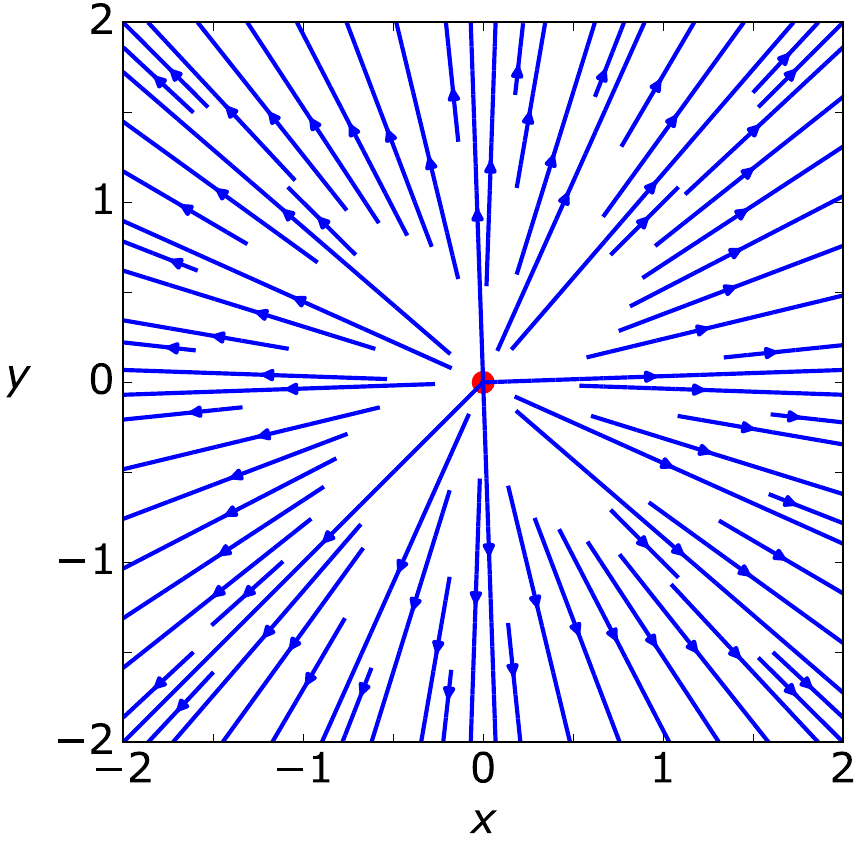}
    \label{fig1b:StreamlineSource}
  }
  \subfigure[]{
    \includegraphics[width=0.3\textwidth]{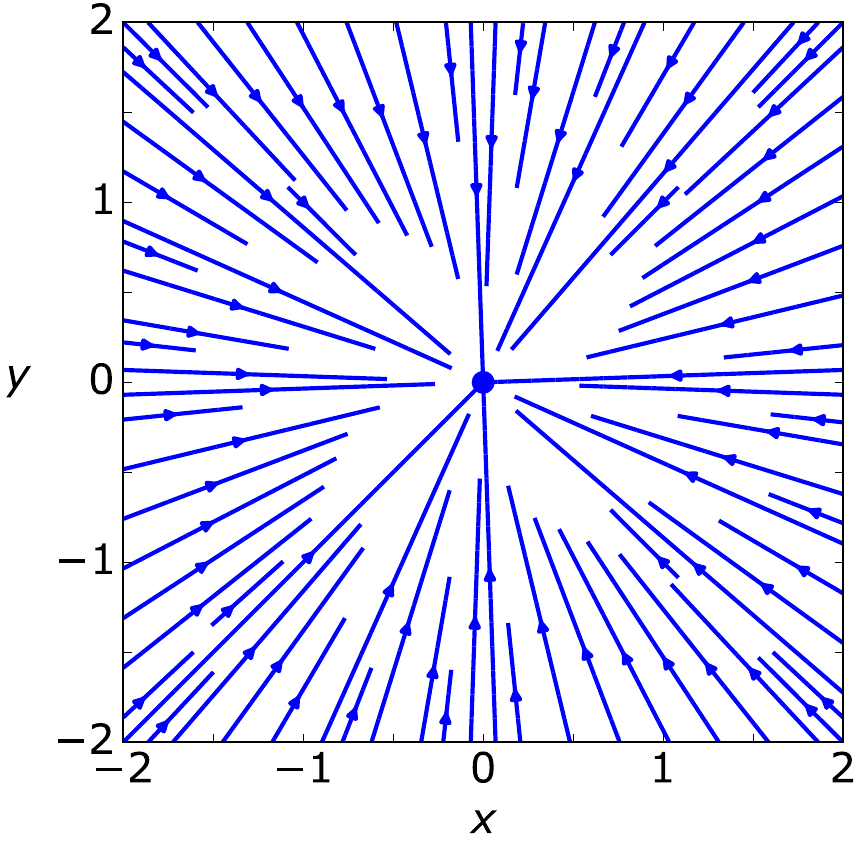}
    \label{fig1c:StreamlineSink}
  }
  \caption{\label{fig1:Streamplots} (a)~Streamlines in a plane containing the swimmer. The swimmer is oriented along the $z$-axis. The source is highlighted in red and the sink in blue. (b)~ Cross-sectional view of the streamlines in the plane $z = 0.5$. (c)~ Cross-sectional view of the streamlines in the plane $z = -0.5$.}
\end{figure}

The flow induced by the motion of a body can be expressed in a multipole expansion. For a body of constant volume, the leading-order term is a dipole \cite[Ch. 4.7]{eldredge2019mathematical}, otherwise called a source doublet \cite[Ch. 6.4]{batchelor2000introduction}. This motivates modeling a swimmer as a source-sink pair of equal strength separated by a fixed distance $\ell$, which yields a dipole flow at large distances while also endowing a length scale absent from a doublet. Effectively, our model represents the far field of a swimmer. As will become clear later, the separation between the source and sink unlocks non-trivial rotational dynamics in our model swimmers. The streamlines in a plane containing the source-sink pair are shown in Figure~\ref{fig1a:StreamlinesPlanar}. From a physical standpoint, the source represents the fore of a swimmer, which pushes fluid away as the swimmer moves forward. Conversely, the sink represents the rear of a swimmer, where fluid must rush in to fill the void left by the rear as the swimmer moves forward. 
To validate the source-sink pair as an effective representation of a swimmer, we compare the flow field generated by our model to that generated by a tuna, as calculated in the simulations of \citet{zhu2002three}. In their work, \citeauthor{zhu2002three} employed a three-dimensional panel method to simulate both fish and fish-like robots in an inviscid flow, which agreed reasonably with experimental measurements. The streamlines produced by the source-sink pair are shown in three different planes in Figure~\ref{fig1:Streamplots}. The qualitative agreement with the streamlines of the higher-fidelity tuna simulations is excellent outside of the wake region (cf. Figure 5 in \cite{zhu2002three}). Even very close to the body (within a small fraction of the length of the body) the qualitative agreement is excellent. The streamlines in the wake region differ, which is to be expected since our model does not include vortex shedding. Despite this limitation, our approach should reasonably model the far-field hydrodynamic interactions between swimmers. We return to a discussion of the limitations of the model later.

\begin{figure}
    \centering
    \includegraphics[width=0.35\textwidth]{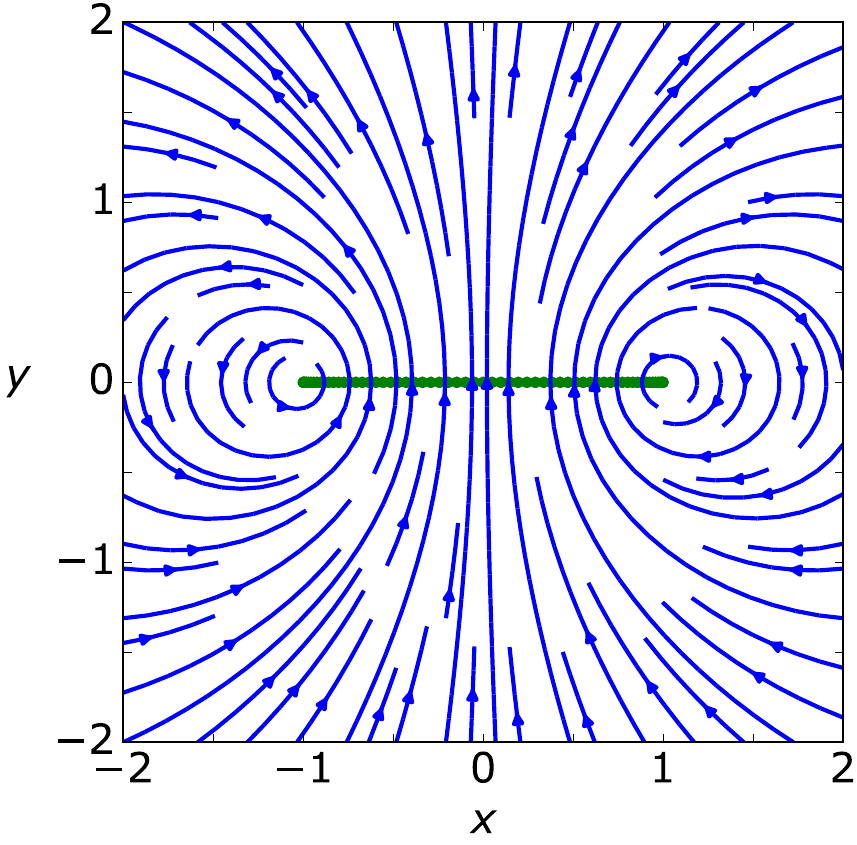}
    \caption{\label{fig2:StreamlinesVortex} Streamlines in the plane of rotational symmetry for a vortex ring.}    
\end{figure}

Alternatively, one could model the swimmer as a thin vortex ring of fixed shape, whose streamlines are shown in Figure~\ref{fig2:StreamlinesVortex}. Although the flow induced by the vortex ring is quite different from that induced by the source-sink pair close to the singularities, the far-field flows are both dipolar and are therefore representative of what one would expect from a swimmer. In prior two-dimensional work, swimmers have been modeled as counter-rotating pairs of point vortices \cite{tchieu2012finite, kanso2014dipole, gazzola2016learning, filella2018model, porfiri2022hydrodynamic}, analogous to a thin vortex ring in three dimensions. 

We explored modeling a swimmer as a thin three-dimensional vortex ring. In our model, the ring was discretized into a set of evenly spaced material points, induced velocities were evaluated using the Biot-Savart law as described in \cite{kimura2017scaling}, the cut-off method was employed to de-singularize induced velocity calculations \cite[Ch. 11.1]{saffman_1993}, and we developed a numerical procedure to enforce that the shape of the vortex ring not change in time. Besides being more complex and computationally expensive than the source-sink pair model, we believe that the vortex ring model is less representative of typical swimmers for reasons that will be clarified in the ensuing section. Thus, the work described here focuses on the source-sink pair model.

\subsection{Dynamics of freely moving swimmers}

At a displacement $\mathbf{r}$ away from a source, the velocity induced by it is
\begin{equation}
  \label{eq:ind}
    \mathbf{u} = \frac{\sigma}{4 \pi} \frac{\mathbf{r}}{r^3},
\end{equation}
where $\sigma \geq 0$ is the volumetric flow rate, and $r = \vert \mathbf{r} \vert$. The velocity induced by a sink is the negative of the above expression. 

The strength of the source and sink that comprise the model swimmer clearly should be related to the speed of the swimmer. We take the swimmer's speed $U$ to be equal in magnitude to the velocity that the source (sink) induces on the sink (source), $U = \frac{\sigma}{4\pi \ell^2}$, but opposite in direction in order to be consistent with the physical picture that the source corresponds to the fore of the body and the sink to the rear. An isolated swimmer moves in the direction parallel to its body at a speed $U$, and we refer to its velocity as the self-propelled velocity. 

When more than one swimmer is present, they will mutually affect each other's motions. To determine the hydrodynamic interactions, we treat each swimmer as a dumbbell comprising two beads connected by a rigid rod, the beads respectively co-located with the source and sink. This dumbbell representation is common in the microswimmer literature \cite{hernandez2005transport, hernandez2007fast, underhill2008diffusion, hernandez2009dynamics, ryan2011viscosity}. In this representation, all the hydrodynamic forces are concentrated on the beads, and the rigid rod serves to keep the length of the dumbbell fixed.

Consider one bead of a swimmer. Let $\mathbf{R}(t)$ be its position, $\mathbf{V}(t)$ be its velocity, and $\mathbf{u}_0(\mathbf{x},t)$ be the velocity field induced by the other bead and swimmers in the absence of the bead under consideration. For an incompressible, irrotational, and inviscid flow, the force on the bead is
\begin{equation}
  \mathbf{F} = \rho \mathcal{V} \left[(1 + C_M)\frac{D\mathbf{u}_0}{Dt}\Big|_{\mathbf{x}=\mathbf{R}} - C_M \frac{\text{d}\mathbf{V}}{\text{d}t}\right],
\end{equation}
where $\rho$ is the density of the fluid, $\mathcal{V}$ is the volume of the bead, $C_M$ is the added mass coefficient of the bead, $\frac{D}{Dt}$ is the material derivative, and $\frac{\text{d}}{\text{d}t}$ is the time derivative following the bead \cite{auton1988force}.

Applying Newton's second law to the bead gives
\begin{equation}
  \frac{\text{d}\mathbf{V}}{\text{d}t} = \frac{1 + C_M}{1 + C_M + \Delta \rho/\rho} \frac{D\mathbf{u}_0}{Dt}\Big|_{\mathbf{x}=\mathbf{R}},
\end{equation}
where $\Delta \rho = \rho_s - \rho$ is the difference between the swimmer's density $\rho_s$ and the fluid's density. When $\Delta \rho = 0$, 
\begin{equation}
  \frac{\text{d}\mathbf{V}}{\text{d}t} = \frac{D\mathbf{u}_0}{Dt}\Big|_{\mathbf{x}=\mathbf{R}}.
\end{equation}
That is, the bead moves as a material point if the density of the swimmer is equal to that of the fluid. 

For each swimmer, therefore, we take the velocity of the source to be equal to the sum of the self-propelled velocity and the velocity induced by the sources and sinks of all other swimmers. The velocity of the sink is calculated the same way. Since swimmers have approximately the same density as the surrounding fluid, treating the velocities of the source and sink in this way is reasonable. In contrast, this may be a poor approach for flying animals (depending on their size) since they are typically significantly denser than the air they fly in; the relatively high inertia of a flying animal would lead it to respond much more slowly to its surrounding flow. 

Left unconstrained, the distance between the source and sink would not remain constant, corresponding to a swimmer whose length changes. To maintain a constant distance between the source and the sink, we adopt the following procedure, which is applied to every swimmer. Let $\mathbf{x}_f$ be the position of the source, $\mathbf{x}_b$ the position of the sink, and $\mathbf{x}_c$ the position of the swimmer's center, so that $\mathbf{x}_c = \frac{1}{2}(\mathbf{x}_f + \mathbf{x}_b)$. The swimmer's configuration is completely specified by the position of its center and its orientation, which is given by a unit vector $\mathbf{n}$ that points from the sink to the source; see Figure~\ref{fig3:SingleSwimmer}. We express $\mathbf{x}_f$ and $\mathbf{x}_b$ in terms of $\mathbf{x}_c$, the swimmer's length $\ell$, and the swimmer's orientation: 
\begin{figure}
    \centering
    \includegraphics[width=0.35\textwidth]{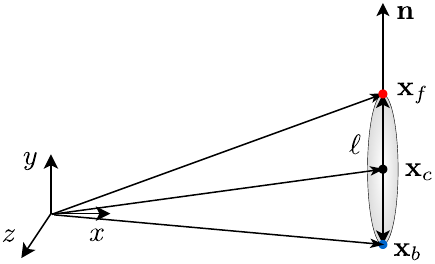}
    \caption{\label{fig3:SingleSwimmer} Representation of a single swimmer.}     
\end{figure}
\begin{equation}
\begin{aligned}
& \mathbf{x}_f=\mathbf{x}_c+\frac{1}{2} \ell \mathbf{n}, \\
& \mathbf{x}_b=\mathbf{x}_c-\frac{1}{2} \ell \mathbf{n}.
\end{aligned}
\end{equation}
The velocities of the source and sink are
\begin{equation}
\label{eq:vel}
\begin{aligned}
& \d{\mathbf{x}_f}{t}=\mathbf{v}_f+\lambda \mathbf{n}, \\
& \d{\mathbf{x}_b}{t} =\mathbf{v}_b-\lambda \mathbf{n},
\end{aligned}
\end{equation}
where $\mathbf{v}_f $ and $\mathbf{v}_b$ respectively correspond to the unconstrained velocities of the source and sink due to the self-propelled velocity and interactions with other swimmers. The additional term $\lambda \mathbf{n}$, in which $\lambda$ may be thought of as a Lagrange multiplier, corresponds to an attractive/repulsive velocity that ensures that the length of the swimmer remains constant: $\d{\ell}{t} = 0$. Adding this term is equivalent to a force that acts on the source and the sink to either push them toward each other or pull them apart. Note that this term acts in the direction of $\mathbf{n}$ and does not produce any net translation or rotation of the swimmer.

To find $\lambda$, take the time derivative of $\mathbf{x}_f - \mathbf{x}_b = \ell \mathbf{n}$ to arrive at
\begin{equation}
  \label{eq:vel2}
 \mathbf{v}_f - \mathbf{v}_b +2 \lambda \mathbf{n} = \d{\ell}{t} \mathbf{n} + \ell \d{\mathbf{n}}{t}.
\end{equation}
 Taking the dot product with $\mathbf{n}$ and recalling that $\d{\mathbf{n}}{t} $ has no component in the direction of $\mathbf{n}$ yields
\begin{equation}
 (\mathbf{v}_f - \mathbf{v}_b) \cdot \mathbf{n} +2 \lambda = \d{\ell}{t} = 0
 \implies \lambda = -\frac{1}{2}  (\mathbf{v}_f - \mathbf{v}_b) \cdot \mathbf{n}. 
\end{equation}

With $\lambda$ in hand, we can find equations for the evolution of the swimmer's position $\mathbf{x}_c$ and orientation $\mathbf{n}$. From~\eqref{eq:vel}, the swimmer translates according to
\begin{equation}
\d{\mathbf{x}_c}{t} = \frac{1}{2}\left(\mathbf{v}_f +\mathbf{v}_b \right),
\label{RHS center}
\end{equation}
and from~\eqref{eq:vel2}, its orientation evolves according to
 \begin{equation}
\d{\mathbf{n}}{t}  = \frac{1}{\ell}\left(\mathbf{v}_f - \mathbf{v}_b +2 \lambda \mathbf{n}\right) = \frac{1}{\ell}\left\{\mathbf{v}_f - \mathbf{v}_b - \left[\left(\mathbf{v}_f - \mathbf{v}_b\right) \cdot \mathbf{n}\right] \mathbf{n}\right\}.
\label{RHS normal}
\end{equation}
The swimmer translates at the sum of the self-propelled velocity and the average of the induced velocities at its source and sink due to other swimmers. Thus, the swimmer's body acts to spatially filter the velocity field induced by all other swimmers. 

\begin{figure}
  \centering
  \subfigure[]{
    \includegraphics[width=0.52\textwidth]{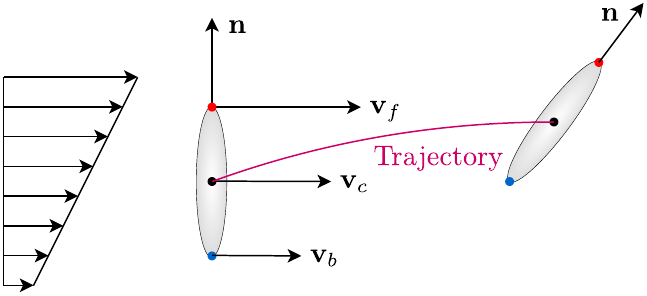}
    \label{fig4a:VeloGrad}
  }\hspace{0.5in}
  \subfigure[]{
    \includegraphics[width=0.14\textwidth]{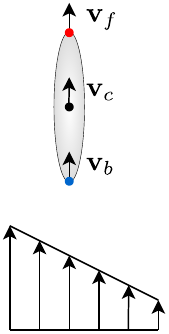}
    \label{fig4b:VeloGradPer}
  }  
  \caption{\label{fig4:VelocityGradients} Rotational dynamics of the source-sink pair model. (a)~ A gradient along the body of the crossflow results in rotation. (b)~ A gradient across the body of the body-parallel velocity does not result in rotation.}
\end{figure}

\begin{figure}
  \centering
  \subfigure[]{
    \includegraphics[width=0.27\textwidth]{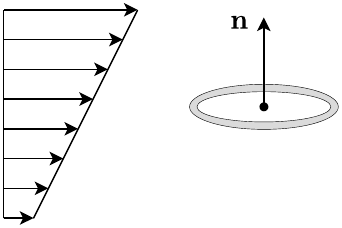}
  }\hspace{1in}
  \subfigure[]{
    \includegraphics[width=0.21\textwidth]{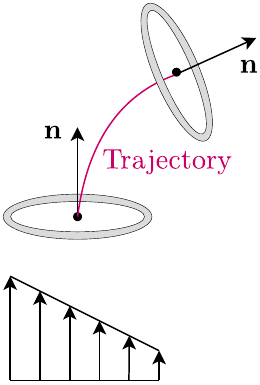}

  }
  \caption{\label{fig6:VelocityGradientsVortex} Rotational dynamics of the vortex ring model. (a)~ A gradient along the body of the crossflow does not result in rotation. (b)~ A gradient across the body of the body-parallel velocity results in rotation.}
\end{figure}

The orientation dynamics are rather interesting. Remark that the swimmer can rotate despite the flow being irrotational. This is possible because of the finite length of the swimmer and the non-zero rate of strain in the flow. In~\eqref{RHS normal}, the term inside of the braces is the component of $\mathbf{v}_f - \mathbf{v}_b$ perpendicular to $\mathbf{n}$, i.e., the component in the plane normal to the swimmer's body. The orientation dynamics are thus given by the gradient along the unit normal of the component of the velocity in the plane normal to the swimmer's body. In other words, the swimmer samples the gradient of the crossflow along its body, leading to rotation. A schematic explanation is illustrated in Figure~\ref{fig4:VelocityGradients}. This is how a short, inextensible material line would deform in a flow. For a swimmer modeled as a thin vortex ring, it would sample the gradient of the body-parallel velocity in the plane of the vortex ring (i.e., the gradient \emph{across} the swimmer's body rather than along it), which leads to qualitatively different dynamics. The rotational dynamics are illustrated in Figure~\ref{fig6:VelocityGradientsVortex}, and should be contrasted with those of the source-sink pair in Figure~\ref{fig4:VelocityGradients}. For swimmers that are long and thin, the source-sink pair model is reasonable since the orientation dynamics of such swimmers would be dominated by the crossflow velocity gradient along the length of their bodies. Conversely, the orientation dynamics of short and fat swimmers would be dominated by the velocity gradient across their bodies, making the vortex ring model appropriate in that case. Since most swimmers are long and thin, we use the source-sink pair model. 

A dynamical system is formed by concatenating the position and unit normal of each swimmer into one state vector and evolving it forward in time according to~\eqref{RHS center} and~\eqref{RHS normal}. The dynamics are nonlinear, so we numerically evolve the system forward in time using the fourth-order Runge-Kutta method. At each step, we re-normalize the unit normal of every swimmer to have unit length since this is not maintained by the Runge-Kutta method. 

The preceding discussion is unchanged for two-dimensional swimmers, except that the velocity induced by a source in~\eqref{eq:ind} is $\mathbf{u} = \frac{\sigma}{2\pi} \frac{\mathbf{r}}{r^2}$, where $\sigma$ is the volumetric flow rate per unit depth in two dimensions. In this case, the self-propelled velocity is $U = \frac{\sigma}{2\pi \ell}$.

\section{\label{sec3:interp}Passive dynamics of a pair}
We present numerical examples of the three-dimensional dynamic interactions between a pair of identical swimmers. Such pairwise interactions can be thought of as the elemental interactions in a large school. To facilitate comparison with two-dimensional models, we focus on planar dynamics, where the swimmers' bodies are co-planar initially and, therefore, for all time. This also allows us to fully characterize the phase space. We also include examples of non-planar dynamics. In what follows, lengths have been non-dimensionalized by the swimmer's length $\ell$ and velocities have been non-dimensionalized by the self-propelled speed $U$. 
 
\subsection{Symmetry reduction}
\begin{figure}
    \centering
    \includegraphics[width=0.2\textwidth]{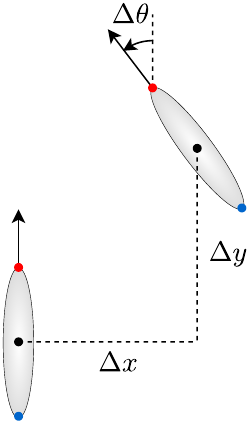}
    \caption{\label{fig6:Symmetry} Symmetry-reduced configuration of a co-planar pair of swimmers.}
\end{figure}

With two swimmers, the dimension of the state space is 12, with six variables for each swimmer (three for the position of its center, and three for the unit normal describing its orientation). For the planar motion of two swimmers, however, leveraging symmetry allows us to describe the motion of the pair using only three variables. Specifically, due to a translational symmetry, only the separation between the swimmers' centers (rather than their absolute positions) affects the dynamics. Similarly, a rotational symmetry makes it so that only the relative angle between the swimmers (rather than their absolute orientations) affects the dynamics. Without loss of generality, we place the swimmers in the $x$-$y$ plane, with one swimmer initially oriented in the positive $y$-direction, separated from each other by $(\Delta x, \Delta y)$, and with a relative angle $\Delta \theta$ between them, as in Figure~\ref{fig6:Symmetry}. In this smaller three-dimensional symmetry-reduced phase space, we can fully characterize the possible dynamics. For non-planar dynamics, an additional angle is required to describe the dynamics. 

\subsection{Planar dynamics of the three-dimensional model}
\begin{figure}
    \centering
    \includegraphics[width=0.5\textwidth]{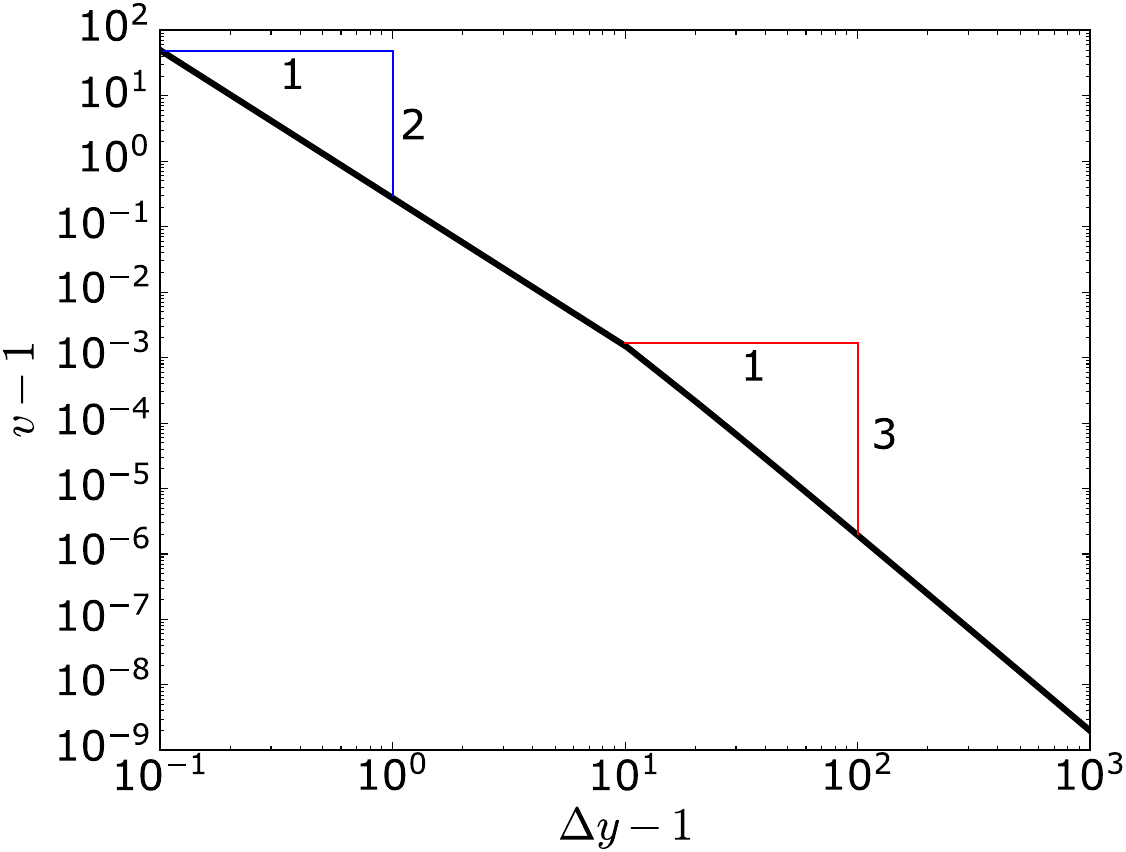}
    \caption{\label{fig7:Asymptotic} Swimmers' speed as a function of their separation for a co-linear configuration for the three-dimensional model.}
\end{figure}

Several qualitatively different sets of dynamics are possible, some simple and expected, others complex and unexpected. The simplest dynamics occur when the swimmers are co-linear ($\Delta x = 0$ and $\Delta \theta \in \{0, \pi \}$). In this case, the swimmers remain co-linear for all time. When the swimmers have the same initial orientation ($\Delta \theta = 0$) and do not intersect ($\vert \Delta y \vert > 1$), they move at equal steady speeds, causing their separation $\Delta y$ to remain constant, no matter its value. Their speed depends on their separation as $v = 1 + \frac{2 \vert \Delta y \vert}{(\Delta y^2 - 1)^2}$. The speed increases monotonically as the swimmers are positioned closer to each other, reflecting a drafting phenomenon. The speed diverges as $\frac{1}{2( \vert \Delta y \vert - 1)^2}$ as $\vert \Delta y \vert \rightarrow 1^+$, and approaches the self-propelled speed as $\frac{2}{\vert \Delta y \vert^3}$ as $\vert \Delta y \vert \rightarrow \infty$, as shown in Figure~\ref{fig7:Asymptotic}. The divergence in speed when the swimmers are very close is non-physical; we must keep in mind that we model the far-field hydrodynamic interactions, which may poorly reflect reality when the swimmers are very close. Since the separation remains constant, such in-line motion forms a locus of relative equilibria. 

When the swimmers are co-linear but have opposite orientations ($\Delta \theta = \pi$), they can either face each other ($\Delta y > 1$) or face away from each other ($\Delta y < -1$). When they face each other, they approach each other at a speed $\d{\Delta y}{t} = -2 + \frac{4\Delta y}{(\Delta y^2 - 1)^2}$. Note that there is a stable equilibrium at $\Delta y \approx 1.684$ where the swimmers directly face each other and do not move. At smaller separations the swimmers push each other away, while at larger separations the swimmers approach this equilibrium. When the swimmers face away from each other, they continually separate at a speed $\d{\Delta y}{t} = 2 - \frac{4\Delta y}{(\Delta y^2 - 1)^2}$, quickly approaching their self-propelled speeds as $\vert \Delta y \vert ^{-3}$. 

\begin{figure}
  \centering
  \subfigure[]{
    \includegraphics[width=0.47\textwidth]{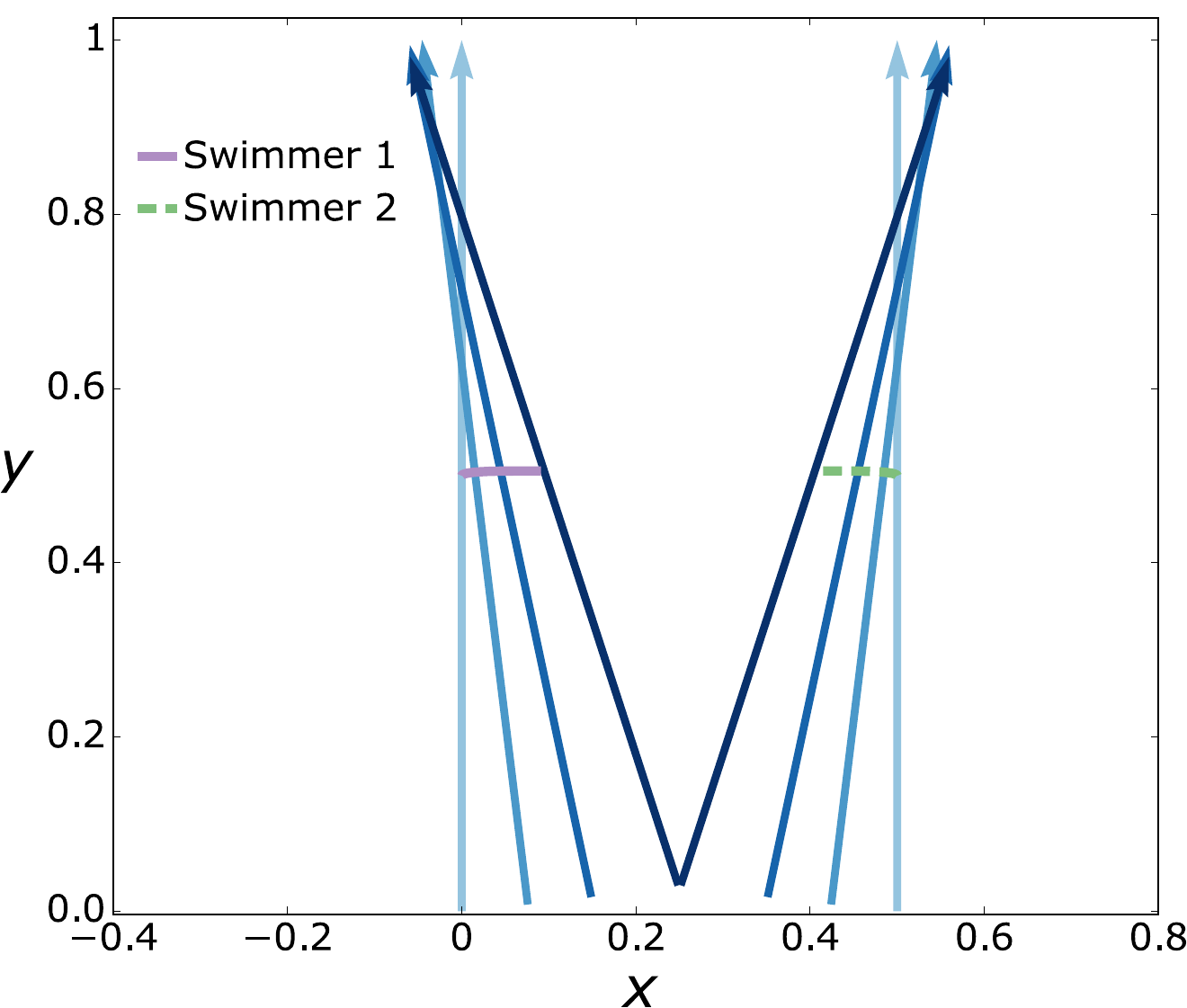}
    \label{fig8a:Collsion}
  }
  \subfigure[]{
    \includegraphics[width=0.47\textwidth]{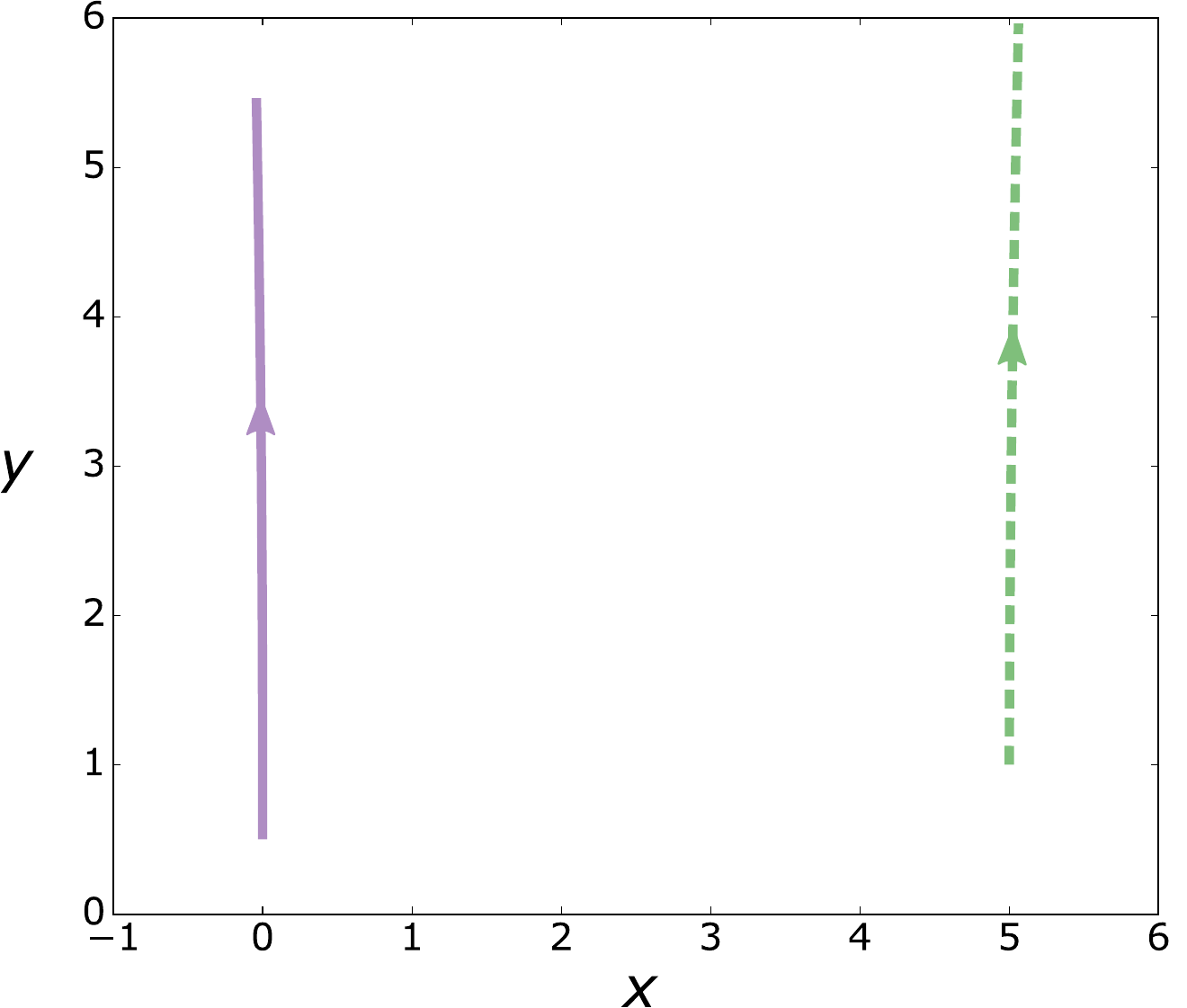}
    \label{fig8b:Diverge}
  }
  \subfigure[]{
    \includegraphics[width=0.47\textwidth]{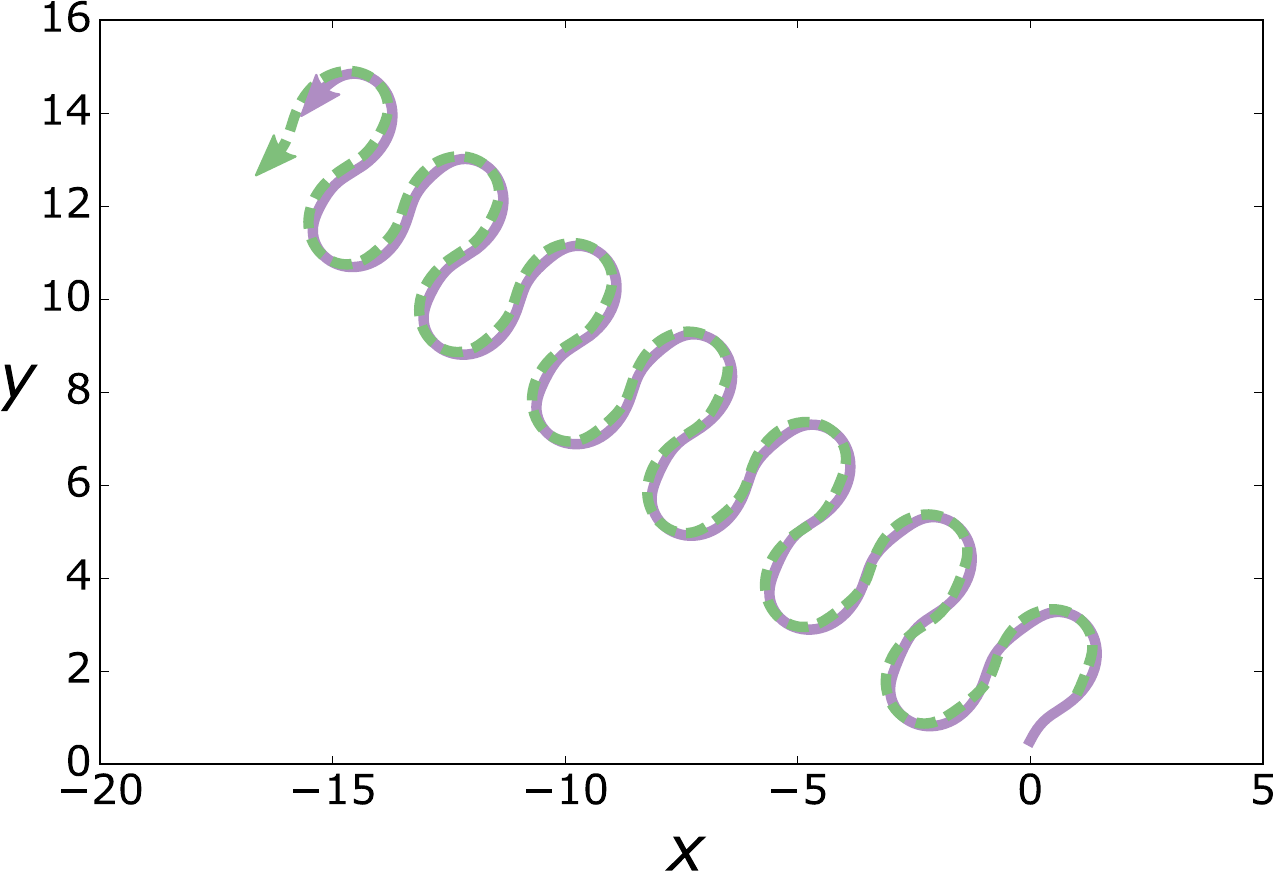}
    \label{fig8c:Braid}
  }
  \subfigure[]{
    \includegraphics[width=0.47\textwidth]{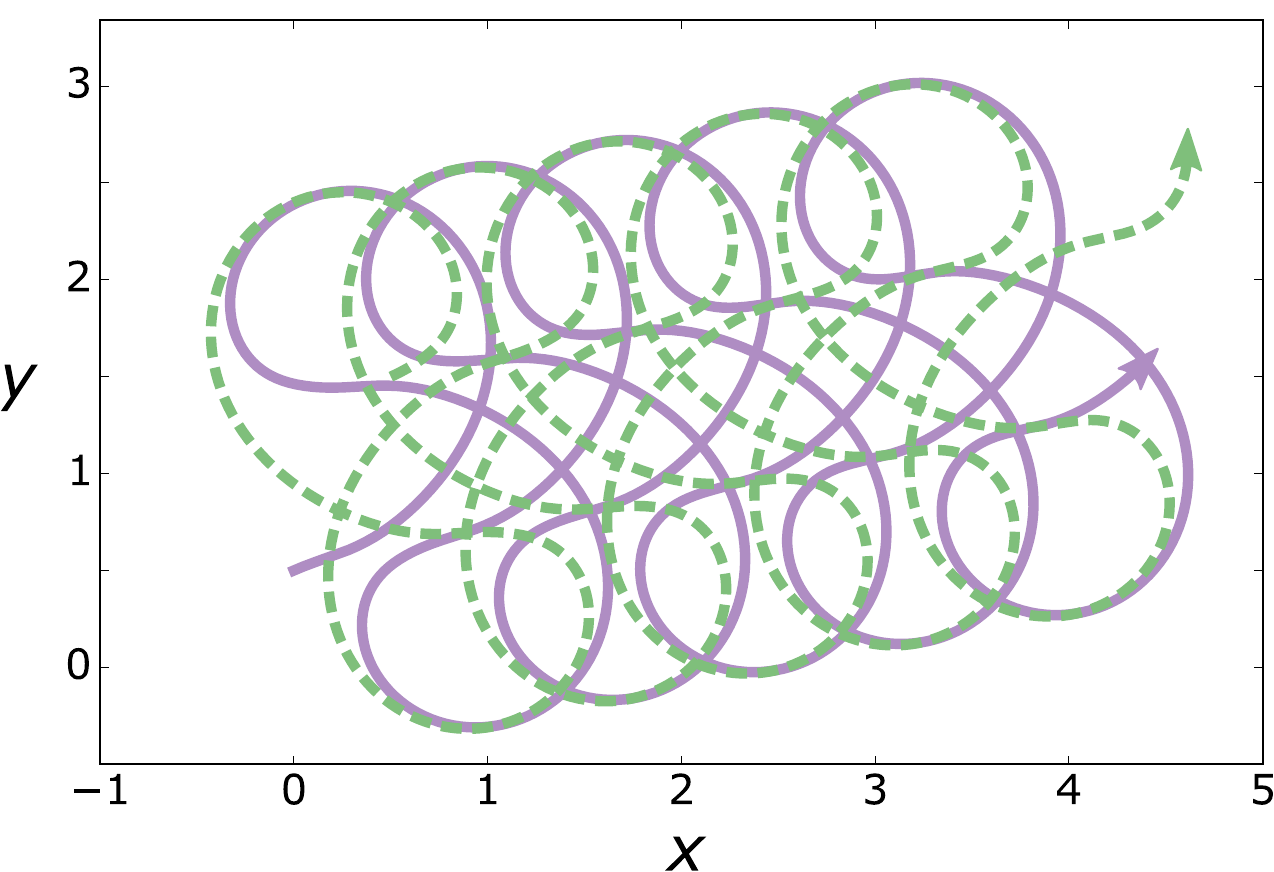}
    \label{fig8d:BraidComplex}
  }
  \caption{\label{fig8:Trajectories} Planar trajectories of two swimmers for different initial conditions with $\Delta \theta = 0$. (a)~ $\Delta \theta = 0$, $\Delta x = 0.5$, $\Delta y = 0$. In (a), the two swimmers' bodies are represented by arrows whose colors represent time (darker is a later time). (b)~ $\Delta \theta = 0$, $\Delta x = 5$, $\Delta y = 0.5$. (c)~ $\Delta \theta = 0$, $\Delta x = 1$, $\Delta y = 1$. (d)~ $\Delta \theta = 0$, $\Delta x = 0.5$, $\Delta y = 1$. }
\end{figure}

For initial configurations where the swimmers are not co-linear, we observe several interesting behaviors. To start, we consider the case when the swimmers have the same orientation ($\Delta \theta = 0$). When the separation between the swimmers is small, they may collide into each other. Such collisions are non-physical---at least for the manner in which they occur under our model, where the swimmers turn away from each other before their rears suck each other in (Figure~\ref{fig8a:Collsion}). At larger separations, the swimmers can diverge from each other (Figure~\ref{fig8b:Diverge}), they can become locked in a simple braid-like motion that is a relative periodic orbit (Figure~\ref{fig8c:Braid}), or they can become locked in a complex braid-like motion that is also a relative periodic orbit (Figure~\ref{fig8d:BraidComplex}). Complex braid-like motions are those where the path of a swimmer intersects itself. For the relative periodic orbits, the relevant symmetry is the translational symmetry. The latter two types of motions correspond to a cohesive pair of swimmers. 

\begin{figure}
    \centering
    \includegraphics[width=0.6\textwidth]{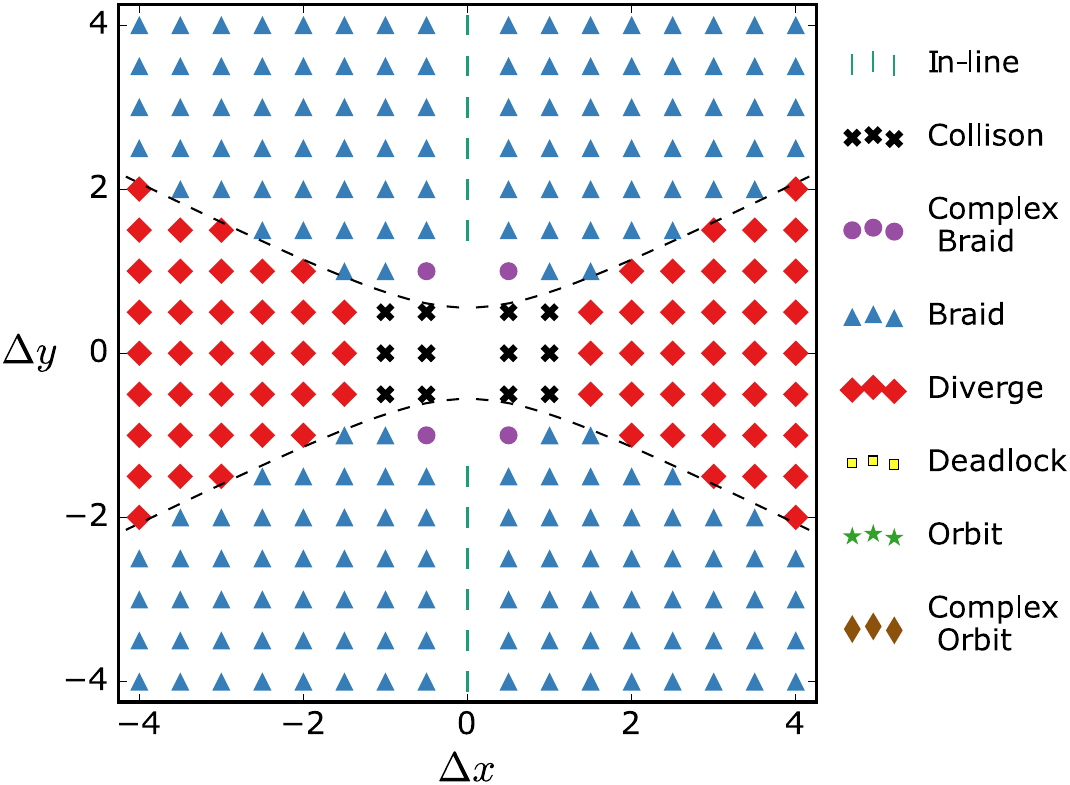}    
    \caption{\label{fig9:phaseDiagram3D} Phase diagram for $\Delta \theta =0$ for three-dimensional swimmers.}
\end{figure}

The possible behaviors for $\Delta \theta = 0$ are summarized in the phase diagram in Figure~\ref{fig9:phaseDiagram3D}, which shows what behaviors are achieved for given initial configurations. There is a clear division between configurations that lead to divergent motions and those that lead to cohesive motions. The boundary between the two is especially of interest to us because we are ultimately interested in whether cohesive groups can be formed by passive hydrodynamic interactions alone. The boundary separates configurations that lead to cohesive groups from those that do not. 

To locate the boundary separating cohesive initial configurations from divergent configurations, we use the fact that a configuration on the boundary is a relative equilibrium of the dynamical system. Moreover, in a boundary configuration, the swimmers do not rotate. Since, without loss of generality, the swimmers' unit normals both point in the positive $y$-direction in their initial configuration, boundary configurations must satisfy $\d{\mathbf{n}}{t} \cdot \mathbf{e}_1 = 0$, where $\mathbf{e}_1$ is the unit vector in the positive $x$-direction. Applying this condition to one of the swimmers yields an implicit equation for the boundary, 
\begin{equation}
    \frac{2 \Delta x}{\left(\Delta x^{2} + \Delta y^{2}\right)^{3/2}} - \frac{\Delta x}{\left[\Delta x^{2} + \left(\Delta y - 1\right)^{2}\right]^{3/2}} - \frac{\Delta x}{\left[\Delta x^{2} + \left(\Delta y + 1\right)^{2}\right]^{3/2}} = 0,
    \label{eqFixed3D}
\end{equation}
which we solve numerically using a root-finding algorithm. At large separations, the boundary is given by $\vert \Delta y \vert \approx \frac{1}{2} \vert \Delta x \vert$. The boundary is drawn in the phase diagram in Figure~\ref{fig9:phaseDiagram3D}. Note the mirror symmetry across both axes. 

\begin{figure}
  \centering
  \subfigure[]{
    \includegraphics[width=0.47\textwidth]{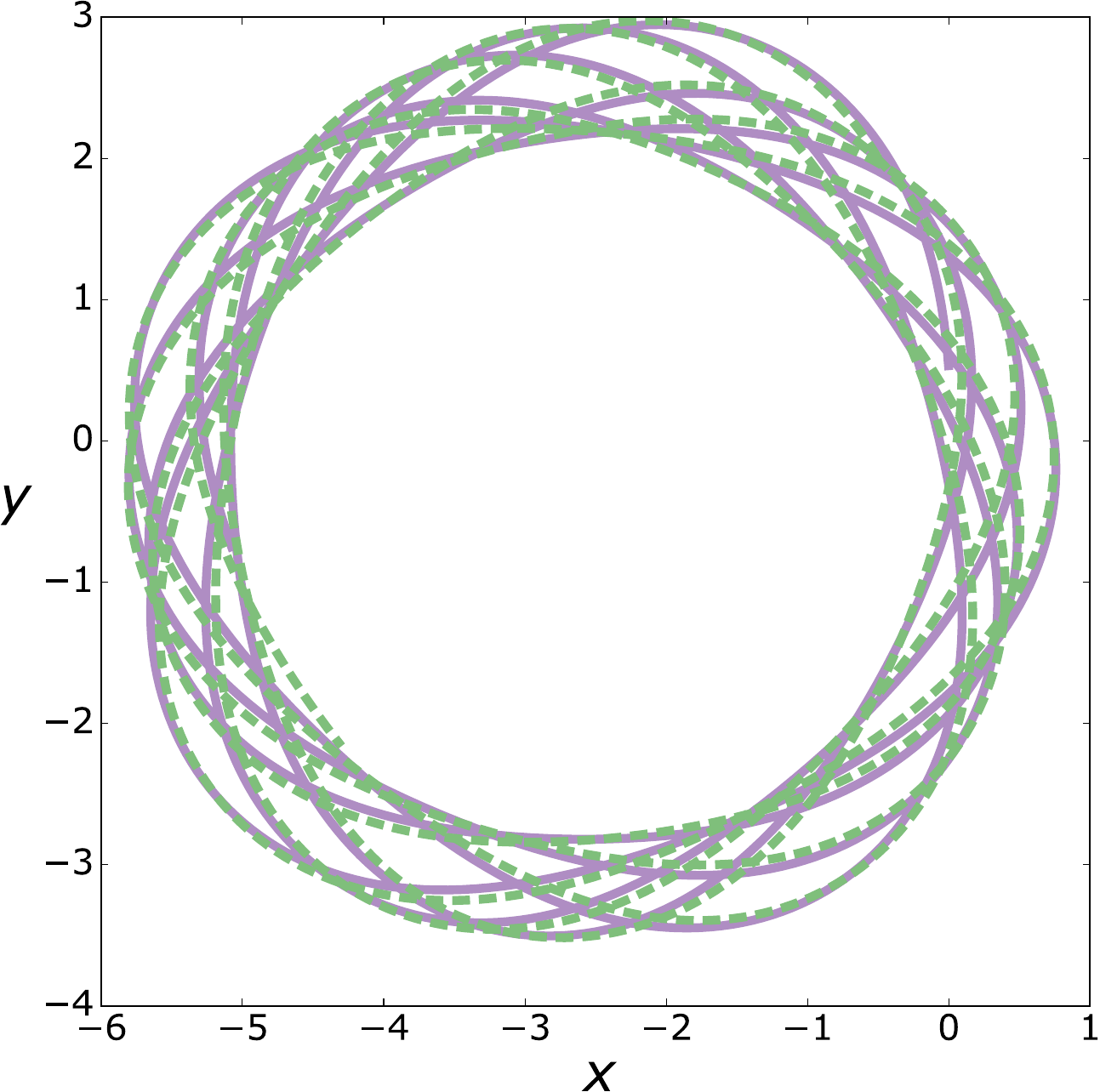}
    \label{fig10a:OrbitComplex1}
  }
  \subfigure[]{
    \includegraphics[width=0.47\textwidth]{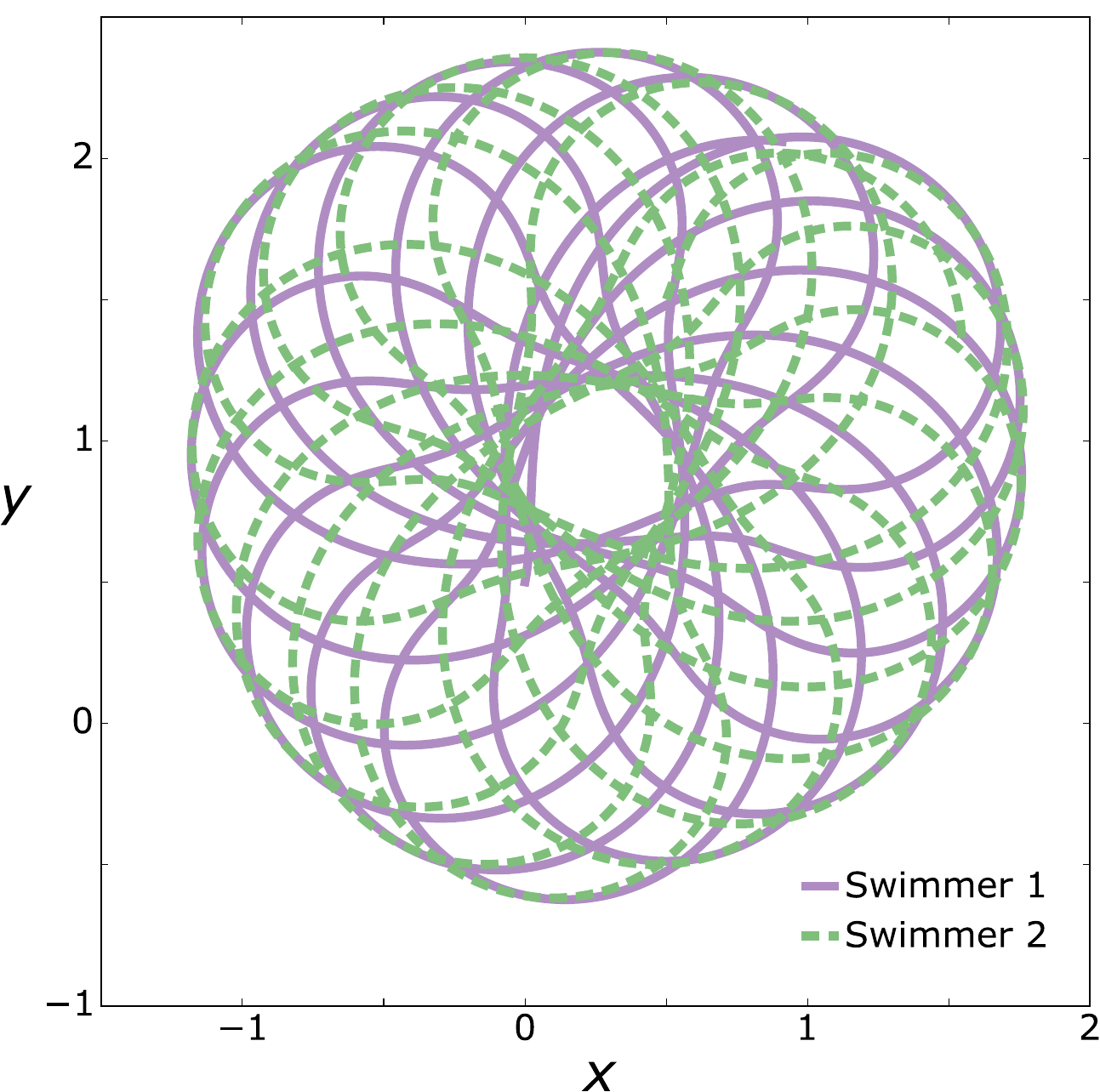}
    \label{fig10b:OrbitComplex}
  }
  \caption{\label{fig11:Trajectories2} Planar trajectories of a pair of swimmers for different initial conditions with $\Delta \theta \neq 0$. ~(a) $\Delta \theta = \frac{\pi}{4}$, $\Delta x = -0.5$, $\Delta y = 1.5$. (b)~$\Delta \theta = \frac{\pi}{3}$, $\Delta x = 0.5$, $\Delta y = 1.0$. }
\end{figure}

For initial configurations with $\Delta \theta \neq 0$, in addition to the classes of motion observed when $\Delta \theta = 0$, we also observe orbital motions (Figure~\ref{fig10a:OrbitComplex1}) and complex orbital motions (Figure~\ref{fig10b:OrbitComplex}). Complex orbital motions are those for which $\d{\theta}{t}$ changes sign for each swimmer, where $\theta$ gives the orientation of a swimmer. Both of these classes of motion correspond to relative periodic orbits, where the relevant symmetry is the rotational symmetry, not the translational symmetry as for the braid-like motions. Without symmetry reduction, the dynamics are quasiperiodic. Notably, swimmers undergoing these motions have no net displacement. 

The possible behaviors for $\Delta \theta \neq 0$ are summarized in the phase diagrams in Appendix~\ref{sec5:AppPD3D}, which show what behaviors are achieved for given initial configurations. In contrast to when $\Delta \theta = 0$, the set of configurations leading to a cohesive group is compact and decreases in size as $\vert \Delta \theta \vert$ increases. In other words, misalignment of the swimmers leads them to diverge from each other. 

\begin{figure}
  \centering
  \subfigure[]{
    \includegraphics[width=0.47\textwidth]{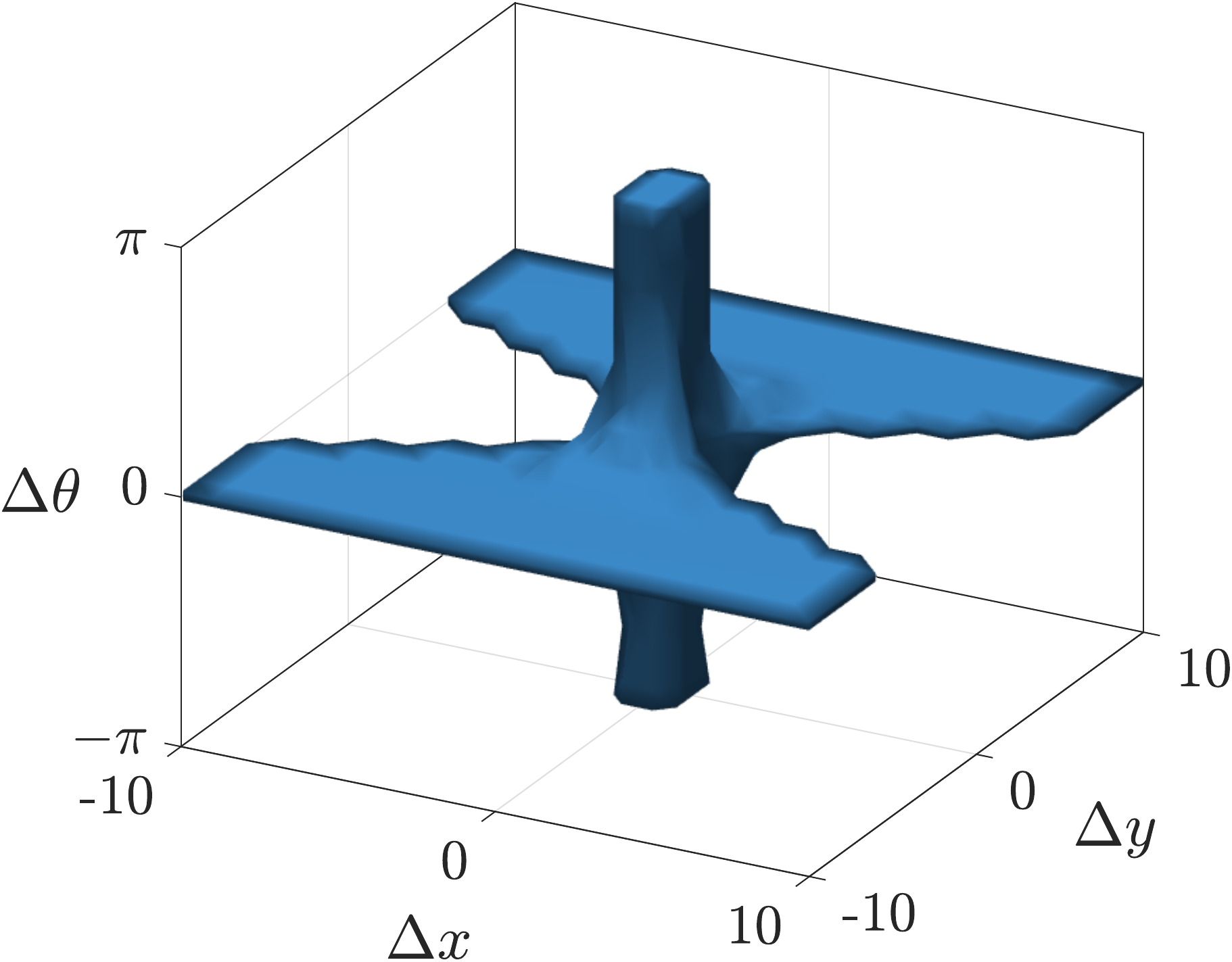}
    \label{fig11a:stateSpace1}
  }
  \subfigure[]{
    \includegraphics[width=0.47\textwidth]{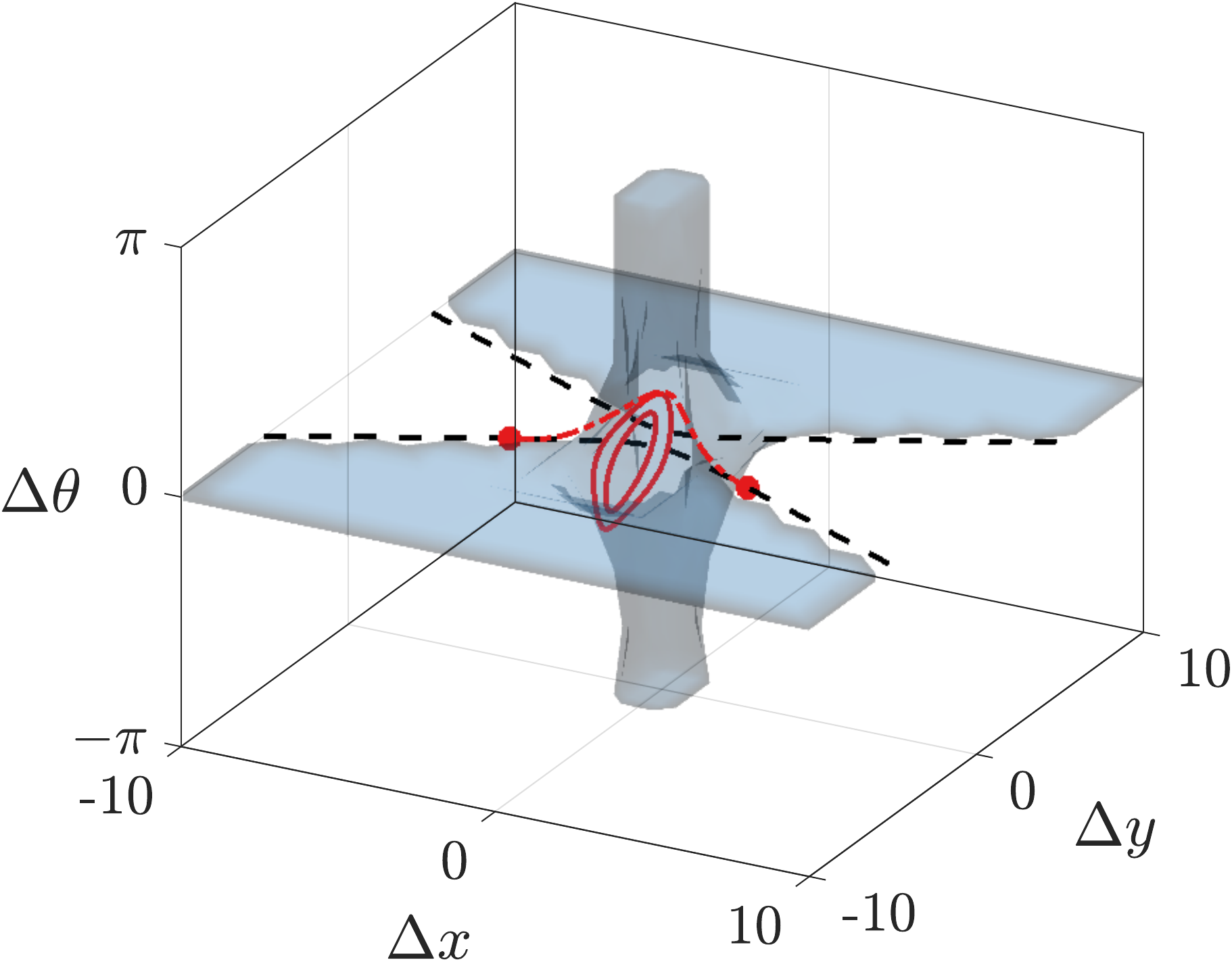}
    \label{fig11b:stateSpace2}
  }
  \caption{(a) Configurations in the symmetry-reduced phase space that lead to a cohesive group. (b)~Dynamical features of importance: relative periodic orbits (solid red); locus of relative equilibria (dashed black); and relative heteroclinic orbit (dash-dotted red) connecting two relative equilibria (red points). }
  \label{fig12:stateSpace}

\end{figure}

We can gain some additional insight from the full three-dimensional symmetry-reduced phase diagram. In Figure~\ref{fig11a:stateSpace1}, we show the volume of configurations that lead to a cohesive group. The central column mostly consists of configurations causing the swimmers to non-physically collide. The volume consists of a continuum of relative periodic orbits, some of which are included for illustration in Figure~\ref{fig11b:stateSpace2}. We have also plotted the boundary separating cohesive configurations from divergent ones for $\Delta \theta = 0$. Recall that the boundary consists of relative equilibria. The bottom and top surfaces of the volume meet at this boundary. These bottom and top surfaces, therefore, contain a continuum of heteroclinic connections in the symmetry-reduced phase space, an example of which is shown in Figure~\ref{fig11b:stateSpace2}. Each relative heteroclinic connection corresponds to a transition from one steadily moving configuration to another. 

\subsection{Planar dynamics of the two-dimensional model}

The phase diagrams for the two-dimensional swimmers are shown in Appendix~\ref{sec6:AppPD2D}. There are no significant qualitative differences between them and the phase diagrams for three-dimensional swimmers. The observations we have made regarding group cohesion for three-dimensional swimmers transfer over to two-dimensional swimmers.

\begin{figure}
    \centering
    \includegraphics[width=0.5\textwidth]{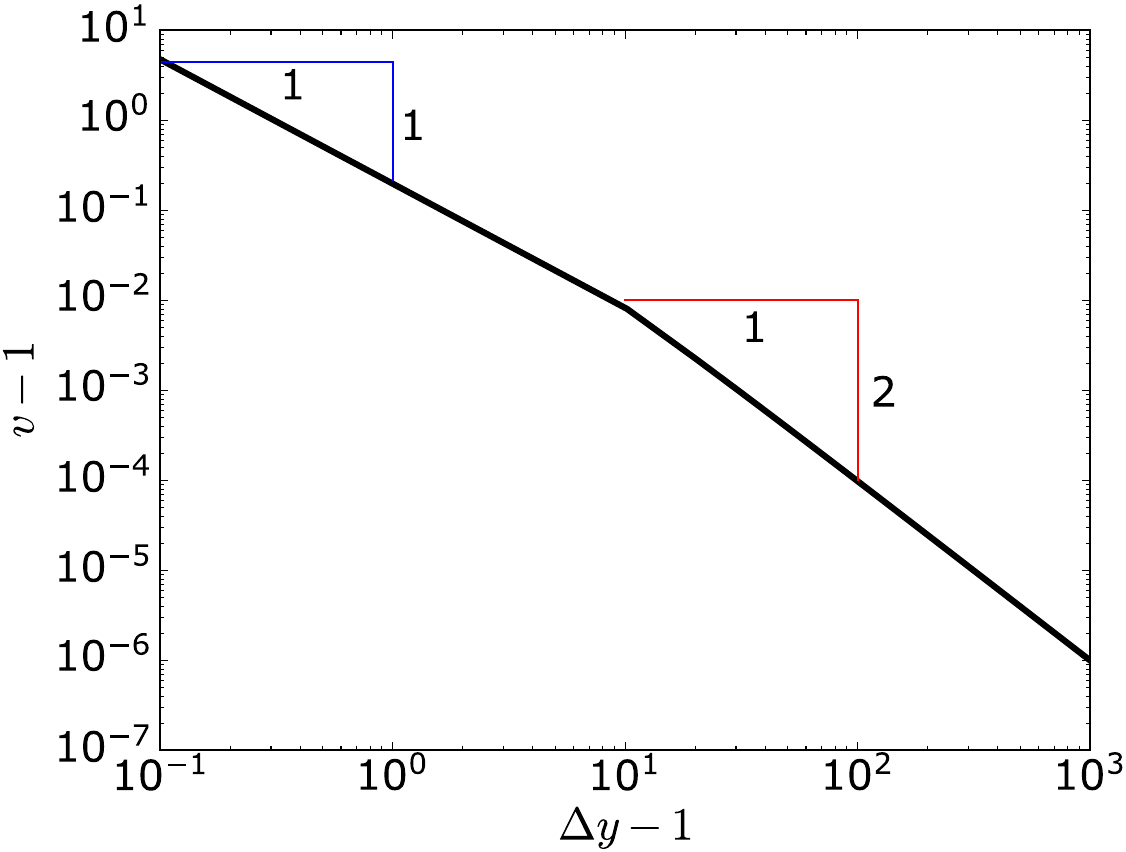}
    \caption{\label{fig12:Asymptotic2D} Swimmers' speed as a function of their separation for a co-linear configuration for the two-dimensional model.}
\end{figure}

There are some quantitative differences worth noting. When the swimmers are co-linear, have the same initial orientation ($\Delta \theta = 0$), and do not intersect ($\vert \Delta y \vert > 1$), they move at equal steady speeds that depend on their separation as $v = 1 + \frac{1}{\Delta y^2 - 1}$. The speed increases monotonically as the swimmers are positioned closer to each other. The speed diverges as $\frac{1}{2( \vert \Delta y \vert - 1)}$ as $\vert \Delta y \vert \rightarrow 1^+$, and approaches the self-propelled speed as $\frac{1}{ \Delta y ^2}$ as $\vert \Delta y \vert \rightarrow \infty$, as shown in Figure~\ref{fig12:Asymptotic2D}. The speed is greater than in the three-dimensional model when $\vert \Delta y \vert > 1 + \sqrt{2}$, but is less for closer separations. When the swimmers are co-linear but have opposite orientations ($\Delta \theta = \pi$), they can either face each other ($\Delta y > 1$) or face away from each other ($\Delta y < -1$). When they face each other, they approach each other at a speed $\d{\Delta y}{t} = -2 + \frac{2}{\Delta y^2 - 1}$. There is a stable equilibrium at $\Delta y = \sqrt{2}$, which is closer than in the three-dimensional model. When the swimmers face away from each other, they continually separate at a speed $\d{\Delta y}{t} = 2 - \frac{2}{\Delta y^2 - 1}$, quickly approaching their self-propelled speeds as $\vert \Delta y \vert ^{-2}$. 

\begin{figure}
    \centering
    \includegraphics[width=0.6\textwidth]{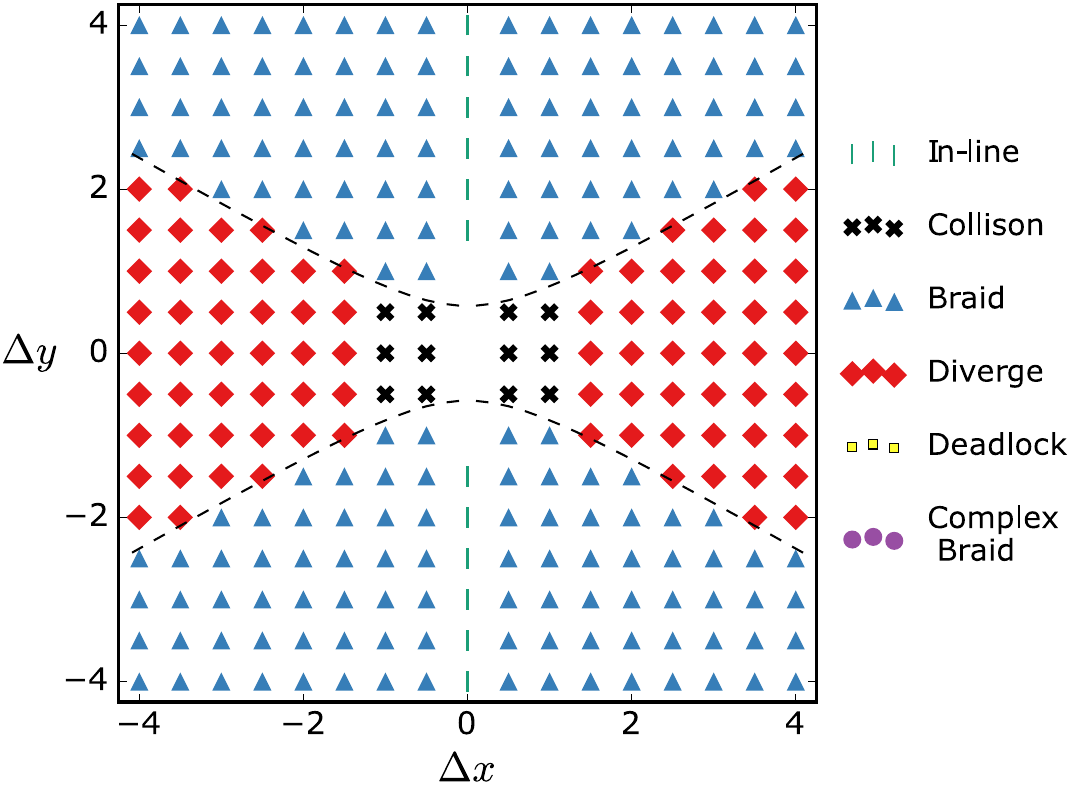}
    \caption{\label{fig13:phaseDiagram2D} Phase diagram for $\Delta \theta =0$ for two-dimensional swimmers.}
\end{figure}

The two-dimensional model also shows a clear boundary in phase space separating cohesive and divergent configurations when $\Delta \theta = 0$, corresponding to a locus of relative equilibria. The phase diagram and boundary are shown in Figure~\ref{fig13:phaseDiagram2D}. The boundary satisfies 
\begin{equation}
    \frac{2\Delta x}{\Delta x^{2} + \Delta y^{2}} - \frac{ \Delta x}{\Delta x^{2} + \left(\Delta y - 1\right)^{2}} - \frac{\Delta x}{\Delta x^{2} + \left(\Delta y + 1\right)^{2}} = 0.
    \label{eqFixed2D}
\end{equation}
At large separations, the boundary is given by $\vert \Delta y \vert \approx \frac{1}{\sqrt{3}}\vert \Delta x \vert$. The set of cohesive configurations is larger for the three-dimensional model. 

Although the two-dimensional swimmers largely have the same behavior as the three-dimensional swimmers, we emphasize that we have only compared the planar dynamics since two-dimensional swimmers are limited to planar dynamics. The three-dimensional model we have developed allows for non-planar dynamics, thereby enabling the study of the dynamics of three-dimensional schools.

\subsection{Non-planar dynamics of the three-dimensional model}

For fully three-dimensional dynamics, an additional angle is required to describe all possible non-redundant initial configurations of a pair of swimmers. Without loss of generality, we place the center of the first swimmer at the origin and direct its unit vector $\mathbf{n}_1$ along the $x$-axis. The center of the second swimmer is placed in the plane $z = 0$, and its unit vector $\mathbf{n}_2 = R_z(\Delta\alpha)R_y(\Delta\beta)\mathbf{n}_1$ is produced by rotating $\mathbf{n}_1$ about the $y$-axis by $\Delta\beta$ and about the $z$-axis by $\Delta\alpha$. The rotation matrices are
\begin{equation}
    R_z(\Delta\alpha) = \begin{bmatrix} \cos\Delta\alpha & -\sin\Delta\alpha & 0 \\
    \sin\Delta\alpha & \cos\Delta\alpha & 0 \\
    0 & 0 & 1 \end{bmatrix}, \quad
    R_y(\Delta\beta) = \begin{bmatrix} \cos\Delta\beta & 0 & \sin\Delta\beta \\
    0 & 1 & 0 \\
    -\sin\Delta\beta & 0 & \cos\Delta\beta \end{bmatrix}.
\end{equation}

\begin{figure}
  \centering
  \subfigure[]{
    \includegraphics[width=0.3\textwidth]{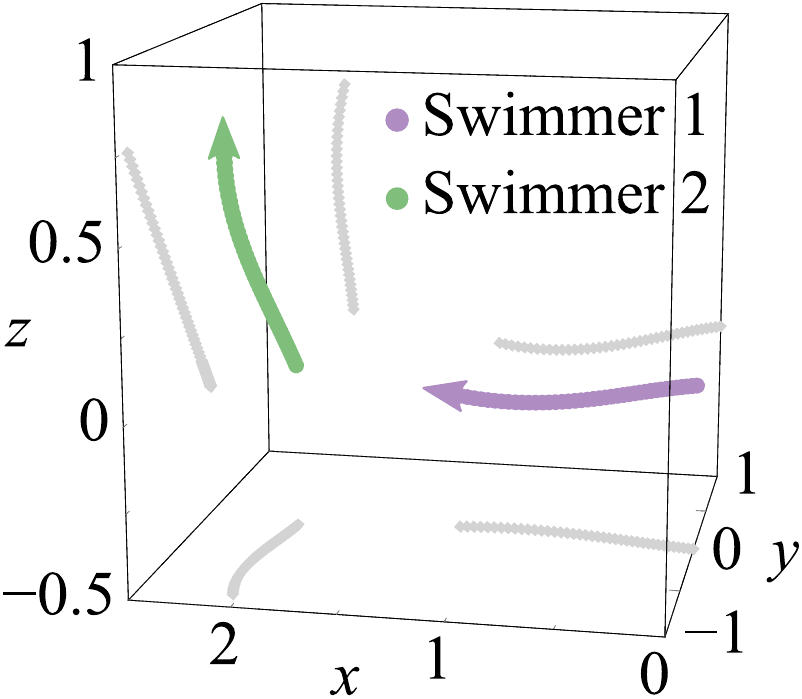}
    \label{fig:3dplot1}
  }
  \subfigure[]{
    \includegraphics[width=0.3\textwidth]{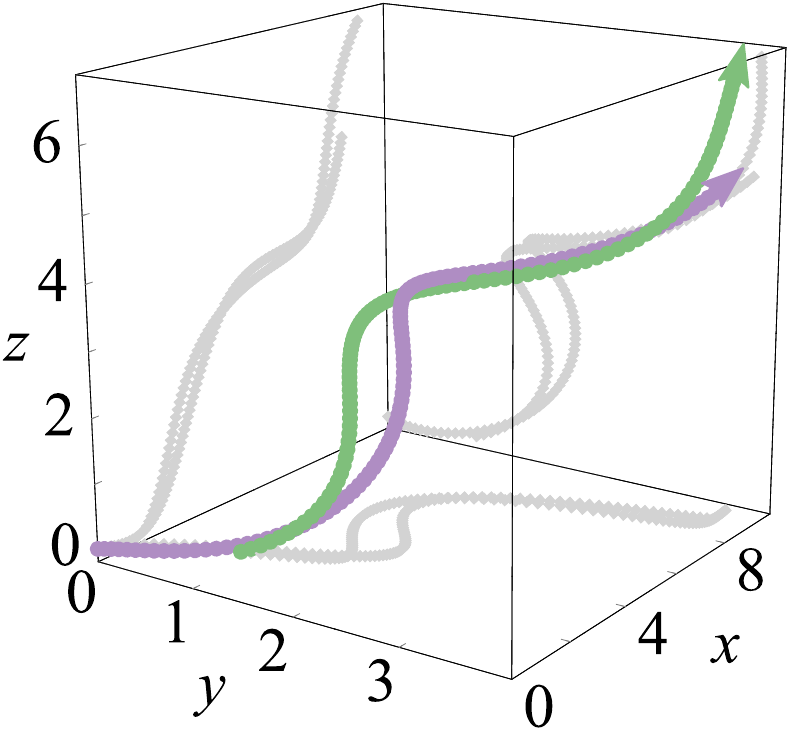}
    \label{fig:3dplot2}
  }
  \subfigure[]{
    \includegraphics[width=0.3\textwidth]{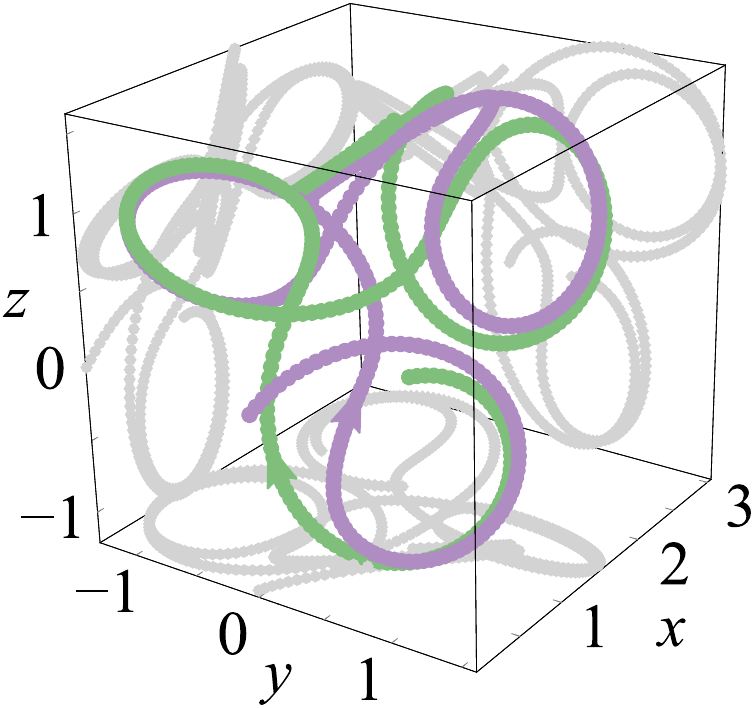}
    \label{fig:3dplot3}
  }
  \caption{Non-planar trajectories of a pair of swimmers for different initial conditions.~(a) $\Delta x = 2$, $\Delta y = 0$, $\Delta z = 0$, $\Delta \alpha = -30$, $\Delta \beta = -15$. (b)~$\Delta x = 1.5$, $\Delta y = 1$, $\Delta z = 0$, $\Delta \alpha = -15$, $\Delta \beta = -15$. (c)~$\Delta x = 1$, $\Delta y = 0.5$, $\Delta z = 0$, $\Delta \alpha = -18$, $\Delta \beta = -18$.}
  \label{fig:3dplot}

\end{figure}

On the basis of the planar results, we expect that the additional degree of freedom present in non-planar dynamics would increase the likelihood that the swimmers diverge since it allows for greater misalignment between the swimmers. An example of divergent motion is shown in Figure~\ref{fig:3dplot1}, where the swimmers' paths have been projected onto three orthogonal planes for clarity. The swimmers initially interact with each other according to their mutually induced velocity fields before separating. Once far enough from each other, the swimmers move along approximately straight-line paths in directions differing from their initial orientations due to the initial interaction. 

Despite the potential for greater misalignment, we nevertheless observe configurations that lead to cohesive motions that are of the same character as in the planar case. In Figure~\ref{fig:3dplot2} we show a three-dimensional braid-like motion analogous to the two-dimensional counterpart in Figure~\ref{fig8c:Braid}. While the planar braid consists of two scales of motion (the oscillation and the mean translation), the non-planar braid consists of three scales: the oscillation apparent in Figure~\ref{fig:3dplot2}, a mean straight-line motion, and a larger-scale oscillation that is approximately six times larger than the oscillation apparent in the plot. In Figure~\ref{fig:3dplot3} we show a configuration that leads to more complex dynamics. For clarity, only a few oscillations are shown. Allowing the motion to evolve further in time reveals net translation of the pair of swimmers. This configuration leads to a three-dimensional complex braid-like motion analogous to the two-dimensional counterpart in Figure~\ref{fig8d:BraidComplex}, although the non-planar complex braid again consists of three scales of motion (the oscillation apparent in Figure~\ref{fig:3dplot3}, a mean straight-line motion, and a larger-scale oscillation), whereas the planar complex braid consists of two scales of motion. 

The non-planar dynamics are apparently very rich, and a detailed investigation will be the subject of future work. Most importantly, passively cohesive groups can exist in the fully three-dimensional problem.

\subsection{Geometric effects}
\begin{figure}
    \centering
    \includegraphics[width=0.6\textwidth]{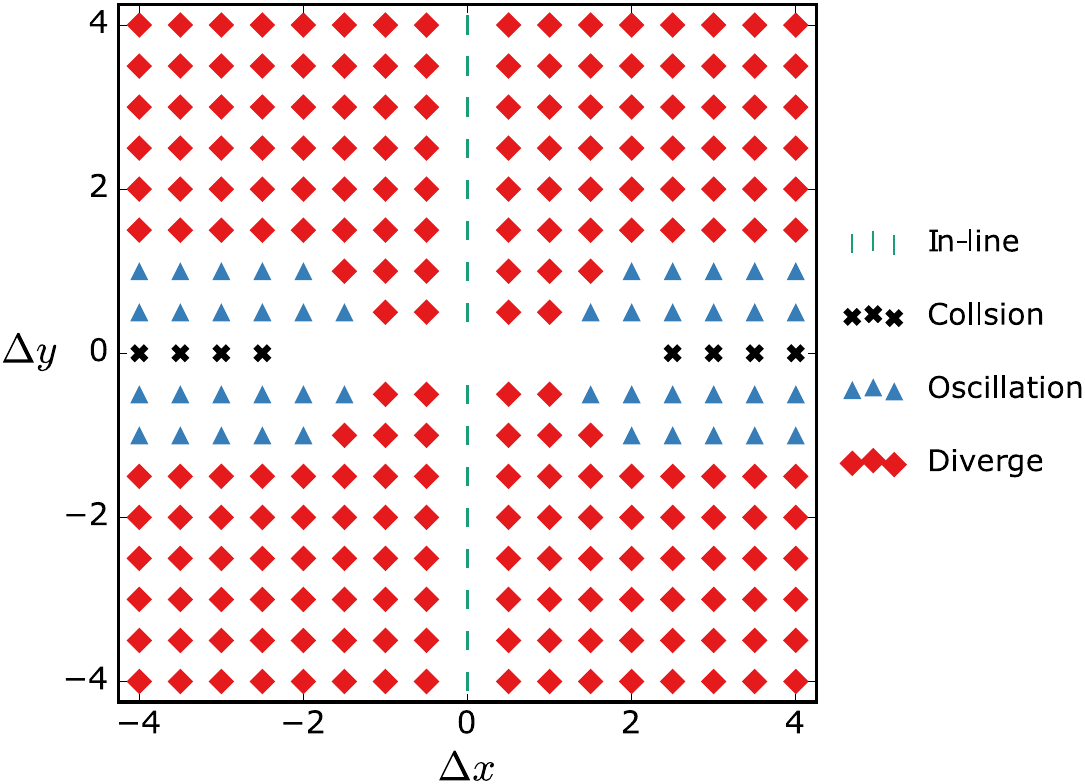}  \caption{\label{fig14:phaseDiagramVortexRing} Phase diagram for $\Delta \theta = 0$ for three-dimensional swimmers modeled as thin vortex rings.}
\end{figure}

Finally, we briefly study the effects of geometry by comparing long and thin swimmers (source-sink pairs) to short and fat swimmers (vortex ring model). The phase diagram for the planar dynamics of two initially aligned ($\Delta \theta = 0$) vortex-ring swimmers is shown in Figure~\ref{fig14:phaseDiagramVortexRing}. The region where the vortex rings overlap is excluded. We observe four types of motions: in-line motion, divergence, collision, and oscillatory motion. Oscillatory motion is qualitatively similar to the braid-like motion in Figure~\ref{fig8:Trajectories}, except that the paths of the swimmers do not intersect, instead following an alternating pattern of convergence and divergence. Although the types of motions are similar to what is observed for the source-sink model, the phase diagram differs substantially. First, the set of configurations leading to a cohesive group is much smaller than for the source-sink model. Second, cohesive configurations have small vertical separations, while for the source-sink model they had large vertical separations. Third, the vortex rings collide only when $\Delta y = 0$; when they collide, they do so by turning head-on towards each other. These qualitative differences in the dynamics of short and fat swimmers compared to long and thin swimmers can be rationalized by appealing to how these two types of swimmers sample the velocity gradient, as was explained in Figures~\ref{fig4:VelocityGradients} and~\ref{fig6:VelocityGradientsVortex}. Interestingly, the dynamics we observe in our vortex ring model are qualitatively similar to the dynamics observed for toroidal swimmers in Stokes flow \cite{huang2017interaction}, despite being at opposite ends of the Reynolds number spectrum. 

\section{\label{sec4:conclu}Conclusion}
The collective motion of inertial swimming animals confers several benefits, but they can be realized only if a cohesive group can be maintained. Group cohesion depends on the dynamics of freely moving swimmers, in which the passive hydrodynamic interactions between individuals play a strong role. This naturally leads one to ask whether a cohesive group can form passively through hydrodynamic interactions alone. 

To understand the passive group dynamics, we developed a simplified model of the three-dimensional hydrodynamic interactions between freely moving inertial swimmers. Our model is based on the leading-order term of the multipole expansion of the velocity field, and is therefore a far-field model. Nevertheless, the model accurately captures the flow created by commonly found long and thin swimmers, at least qualitatively. Of note is the very low computational cost of our model. 

Since the passive dynamics of swimmers with three-dimensional hydrodynamic interactions had not been studied before, we first sought to understand the role of three-dimensionality. This was accomplished by studying the planar dynamics of a pair of swimmers and comparing them to the dynamics of a pair of swimmers with two-dimensional hydrodynamic interactions. We observed many interesting dynamics: braid-like motions, orbiting motions, divergent motions, and others, which were thoroughly characterized by phase diagrams. In state space, we observed relative equilibria, relative periodic orbits, quasiperiodic orbits, and relative heteroclinic connections. Most importantly, we found that certain configurations of the swimmers led to a cohesive group, and that misalignment of the swimmers led them to diverge from each other. 

The two-dimensional swimmers largely have the same planar dynamics as the three-dimensional swimmers. However, our three-dimensional model enables the study of non-planar dynamics and three-dimensional schools. The non-planar dynamics are of the same character as the planar dynamics, although we observe the emergence of multiscale dynamics in the fully three-dimensional problem. Despite the potential for greater misalignment between swimmers, we observe configurations that lead to a passively cohesive group. A detailed investigation of the non-planar problem will be the subject of future work. 

We also studied the effects of geometry, finding that long and thin swimmers have qualitatively different dynamics than short and fat swimmers. The differences are due to the geometry affecting how the velocity gradient is sampled: the rotational dynamics of long and thin swimmers are dominated by the gradient along their bodies of the crossflow, while for short and fat swimmers they are dominated by the gradient across their bodies of the body-parallel flow. 

In principle, models that make no simplifications can be developed. This would amount to direct numerical simulations of schools of freely moving swimmers. The largest such three-dimensional simulation that we are aware of is a recent simulation of a school of 350 swimmers \cite{chatzimanolakis2023high}. This single simulation required parallel computing with adaptive meshing, and significant compute time on a large computing cluster. With a Reynolds number of 4000 (using the flapping speed as the velocity scale), the simulated swimmers correspond to a typical natural swimmer of roughly 1 cm in length \cite{gazzola2014scaling}. This simulation is a feat of scientific computing, but it also underscores that, as a community, we are incredibly far from having the ability to conduct full-fidelity simulations of large groups of large swimmers. The model we have developed in the present work allows us to explore this very interesting space of large groups of inertia-dominated swimmers (at the cost of truncated physics). Ultimately, we aim to provide insights into suspensions of inertial swimmers, as has been done in the viscosity-dominated limit \cite{hernandez2005transport}. \\

This work was supported by the University of Houston Grants to Enhance and Advance Research Program, 000189684.

\bibliography{apssamp.bib}

\begin{thebibliography}{63}%
\makeatletter
\providecommand \@ifxundefined [1]{%
 \@ifx{#1\undefined}
}%
\providecommand \@ifnum [1]{%
 \ifnum #1\expandafter \@firstoftwo
 \else \expandafter \@secondoftwo
 \fi
}%
\providecommand \@ifx [1]{%
 \ifx #1\expandafter \@firstoftwo
 \else \expandafter \@secondoftwo
 \fi
}%
\providecommand \natexlab [1]{#1}%
\providecommand \enquote  [1]{``#1''}%
\providecommand \bibnamefont  [1]{#1}%
\providecommand \bibfnamefont [1]{#1}%
\providecommand \citenamefont [1]{#1}%
\providecommand \href@noop [0]{\@secondoftwo}%
\providecommand \href [0]{\begingroup \@sanitize@url \@href}%
\providecommand \@href[1]{\@@startlink{#1}\@@href}%
\providecommand \@@href[1]{\endgroup#1\@@endlink}%
\providecommand \@sanitize@url [0]{\catcode `\\12\catcode `\$12\catcode
  `\&12\catcode `\#12\catcode `\^12\catcode `\_12\catcode `\%12\relax}%
\providecommand \@@startlink[1]{}%
\providecommand \@@endlink[0]{}%
\providecommand \url  [0]{\begingroup\@sanitize@url \@url }%
\providecommand \@url [1]{\endgroup\@href {#1}{\urlprefix }}%
\providecommand \urlprefix  [0]{URL }%
\providecommand \Eprint [0]{\href }%
\providecommand \doibase [0]{https://doi.org/}%
\providecommand \selectlanguage [0]{\@gobble}%
\providecommand \bibinfo  [0]{\@secondoftwo}%
\providecommand \bibfield  [0]{\@secondoftwo}%
\providecommand \translation [1]{[#1]}%
\providecommand \BibitemOpen [0]{}%
\providecommand \bibitemStop [0]{}%
\providecommand \bibitemNoStop [0]{.\EOS\space}%
\providecommand \EOS [0]{\spacefactor3000\relax}%
\providecommand \BibitemShut  [1]{\csname bibitem#1\endcsname}%
\let\auto@bib@innerbib\@empty
\bibitem [{\citenamefont {Vicsek}\ and\ \citenamefont
  {Zafeiris}(2012)}]{vicsek2012collective}%
  \BibitemOpen
  \bibfield  {author} {\bibinfo {author} {\bibfnamefont {T.}~\bibnamefont
  {Vicsek}}\ and\ \bibinfo {author} {\bibfnamefont {A.}~\bibnamefont
  {Zafeiris}},\ }\bibfield  {title} {\bibinfo {title} {Collective motion},\
  }\href@noop {} {\bibfield  {journal} {\bibinfo  {journal} {Physics Reports}\
  }\textbf {\bibinfo {volume} {517}},\ \bibinfo {pages} {71} (\bibinfo {year}
  {2012})}\BibitemShut {NoStop}%
\bibitem [{\citenamefont {Gautrais}\ \emph {et~al.}(2008)\citenamefont
  {Gautrais}, \citenamefont {Jost},\ and\ \citenamefont
  {Theraulaz}}]{e3c99d1b-f54c-39bd-90da-8cd2a5ff32ba}%
  \BibitemOpen
  \bibfield  {author} {\bibinfo {author} {\bibfnamefont {J.}~\bibnamefont
  {Gautrais}}, \bibinfo {author} {\bibfnamefont {C.}~\bibnamefont {Jost}},\
  and\ \bibinfo {author} {\bibfnamefont {G.}~\bibnamefont {Theraulaz}},\
  }\bibfield  {title} {\bibinfo {title} {Key behavioural factors in a
  self-organised fish school model},\ }\href
  {http://www.jstor.org/stable/23736910} {\bibfield  {journal} {\bibinfo
  {journal} {Annales Zoologici Fennici}\ }\textbf {\bibinfo {volume} {45}},\
  \bibinfo {pages} {415} (\bibinfo {year} {2008})}\BibitemShut {NoStop}%
\bibitem [{\citenamefont {Major}(1978)}]{major1978predator}%
  \BibitemOpen
  \bibfield  {author} {\bibinfo {author} {\bibfnamefont {P.~F.}\ \bibnamefont
  {Major}},\ }\bibfield  {title} {\bibinfo {title} {Predator-prey interactions
  in two schooling fishes, {C}aranx ignobilis and {S}tolephorus purpureus},\
  }\href@noop {} {\bibfield  {journal} {\bibinfo  {journal} {Animal Behaviour}\
  }\textbf {\bibinfo {volume} {26}},\ \bibinfo {pages} {760} (\bibinfo {year}
  {1978})}\BibitemShut {NoStop}%
\bibitem [{\citenamefont {Shaw}(1978)}]{shaw1978schooling}%
  \BibitemOpen
  \bibfield  {author} {\bibinfo {author} {\bibfnamefont {E.}~\bibnamefont
  {Shaw}},\ }\bibfield  {title} {\bibinfo {title} {Schooling fishes},\
  }\href@noop {} {\bibfield  {journal} {\bibinfo  {journal} {American
  Scientist}\ }\textbf {\bibinfo {volume} {66}},\ \bibinfo {pages} {166}
  (\bibinfo {year} {1978})}\BibitemShut {NoStop}%
\bibitem [{\citenamefont {Landeau}\ and\ \citenamefont
  {Terborgh}(1986)}]{landeau1986oddity}%
  \BibitemOpen
  \bibfield  {author} {\bibinfo {author} {\bibfnamefont {L.}~\bibnamefont
  {Landeau}}\ and\ \bibinfo {author} {\bibfnamefont {J.}~\bibnamefont
  {Terborgh}},\ }\bibfield  {title} {\bibinfo {title} {Oddity and the
  `confusion effect' in predation},\ }\href@noop {} {\bibfield  {journal}
  {\bibinfo  {journal} {Animal Behaviour}\ }\textbf {\bibinfo {volume} {34}},\
  \bibinfo {pages} {1372} (\bibinfo {year} {1986})}\BibitemShut {NoStop}%
\bibitem [{\citenamefont {Pitcher}\ \emph {et~al.}(1982)\citenamefont
  {Pitcher}, \citenamefont {Magurran},\ and\ \citenamefont
  {Winfield}}]{pitcher1982fish}%
  \BibitemOpen
  \bibfield  {author} {\bibinfo {author} {\bibfnamefont {T.~J.}\ \bibnamefont
  {Pitcher}}, \bibinfo {author} {\bibfnamefont {A.~E.}\ \bibnamefont
  {Magurran}},\ and\ \bibinfo {author} {\bibfnamefont {I.~J.}\ \bibnamefont
  {Winfield}},\ }\bibfield  {title} {\bibinfo {title} {Fish in larger shoals
  find food faster},\ }\href@noop {} {\bibfield  {journal} {\bibinfo  {journal}
  {Behavioral Ecology and Sociobiology}\ }\textbf {\bibinfo {volume} {10}},\
  \bibinfo {pages} {149} (\bibinfo {year} {1982})}\BibitemShut {NoStop}%
\bibitem [{\citenamefont {Barnes}\ and\ \citenamefont
  {Hughes}(1999)}]{barnes1999introduction}%
  \BibitemOpen
  \bibfield  {author} {\bibinfo {author} {\bibfnamefont {R.~S.~K.}\
  \bibnamefont {Barnes}}\ and\ \bibinfo {author} {\bibfnamefont {R.~N.}\
  \bibnamefont {Hughes}},\ }\href@noop {} {\emph {\bibinfo {title} {An
  introduction to marine ecology}}}\ (\bibinfo  {publisher} {John Wiley \&
  Sons},\ \bibinfo {year} {1999})\BibitemShut {NoStop}%
\bibitem [{\citenamefont {Weihs}(1973)}]{weihs1973hydromechanics}%
  \BibitemOpen
  \bibfield  {author} {\bibinfo {author} {\bibfnamefont {D.}~\bibnamefont
  {Weihs}},\ }\bibfield  {title} {\bibinfo {title} {Hydromechanics of fish
  schooling},\ }\href@noop {} {\bibfield  {journal} {\bibinfo  {journal}
  {Nature}\ }\textbf {\bibinfo {volume} {241}},\ \bibinfo {pages} {290}
  (\bibinfo {year} {1973})}\BibitemShut {NoStop}%
\bibitem [{\citenamefont {Brambilla}\ \emph {et~al.}(2013)\citenamefont
  {Brambilla}, \citenamefont {Ferrante}, \citenamefont {Birattari},\ and\
  \citenamefont {Dorigo}}]{brambilla2013swarm}%
  \BibitemOpen
  \bibfield  {author} {\bibinfo {author} {\bibfnamefont {M.}~\bibnamefont
  {Brambilla}}, \bibinfo {author} {\bibfnamefont {E.}~\bibnamefont {Ferrante}},
  \bibinfo {author} {\bibfnamefont {M.}~\bibnamefont {Birattari}},\ and\
  \bibinfo {author} {\bibfnamefont {M.}~\bibnamefont {Dorigo}},\ }\bibfield
  {title} {\bibinfo {title} {Swarm robotics: a review from the swarm
  engineering perspective},\ }\href@noop {} {\bibfield  {journal} {\bibinfo
  {journal} {Swarm Intelligence}\ }\textbf {\bibinfo {volume} {7}},\ \bibinfo
  {pages} {1} (\bibinfo {year} {2013})}\BibitemShut {NoStop}%
\bibitem [{\citenamefont {Rubenstein}\ \emph {et~al.}(2014)\citenamefont
  {Rubenstein}, \citenamefont {Cornejo},\ and\ \citenamefont
  {Nagpal}}]{rubenstein2014programmable}%
  \BibitemOpen
  \bibfield  {author} {\bibinfo {author} {\bibfnamefont {M.}~\bibnamefont
  {Rubenstein}}, \bibinfo {author} {\bibfnamefont {A.}~\bibnamefont
  {Cornejo}},\ and\ \bibinfo {author} {\bibfnamefont {R.}~\bibnamefont
  {Nagpal}},\ }\bibfield  {title} {\bibinfo {title} {Programmable self-assembly
  in a thousand-robot swarm},\ }\href@noop {} {\bibfield  {journal} {\bibinfo
  {journal} {Science}\ }\textbf {\bibinfo {volume} {345}},\ \bibinfo {pages}
  {795} (\bibinfo {year} {2014})}\BibitemShut {NoStop}%
\bibitem [{\citenamefont {Dorigo}\ \emph {et~al.}(2020)\citenamefont {Dorigo},
  \citenamefont {Theraulaz},\ and\ \citenamefont
  {Trianni}}]{dorigo2020reflections}%
  \BibitemOpen
  \bibfield  {author} {\bibinfo {author} {\bibfnamefont {M.}~\bibnamefont
  {Dorigo}}, \bibinfo {author} {\bibfnamefont {G.}~\bibnamefont {Theraulaz}},\
  and\ \bibinfo {author} {\bibfnamefont {V.}~\bibnamefont {Trianni}},\
  }\bibfield  {title} {\bibinfo {title} {Reflections on the future of swarm
  robotics},\ }\href@noop {} {\bibfield  {journal} {\bibinfo  {journal}
  {Science Robotics}\ }\textbf {\bibinfo {volume} {5}},\ \bibinfo {pages}
  {eabe4385} (\bibinfo {year} {2020})}\BibitemShut {NoStop}%
\bibitem [{\citenamefont {Gordon}(2010)}]{gordon2010ant}%
  \BibitemOpen
  \bibfield  {author} {\bibinfo {author} {\bibfnamefont {D.~M.}\ \bibnamefont
  {Gordon}},\ }\href@noop {} {\emph {\bibinfo {title} {Ant encounters:
  interaction networks and colony behavior}}}\ (\bibinfo  {publisher}
  {Princeton University Press},\ \bibinfo {year} {2010})\BibitemShut {NoStop}%
\bibitem [{\citenamefont {Li}\ \emph {et~al.}(2020)\citenamefont {Li},
  \citenamefont {Nagy}, \citenamefont {Graving}, \citenamefont {Bak-Coleman},
  \citenamefont {Xie},\ and\ \citenamefont {Couzin}}]{li2020vortex}%
  \BibitemOpen
  \bibfield  {author} {\bibinfo {author} {\bibfnamefont {L.}~\bibnamefont
  {Li}}, \bibinfo {author} {\bibfnamefont {M.}~\bibnamefont {Nagy}}, \bibinfo
  {author} {\bibfnamefont {J.~M.}\ \bibnamefont {Graving}}, \bibinfo {author}
  {\bibfnamefont {J.}~\bibnamefont {Bak-Coleman}}, \bibinfo {author}
  {\bibfnamefont {G.}~\bibnamefont {Xie}},\ and\ \bibinfo {author}
  {\bibfnamefont {I.~D.}\ \bibnamefont {Couzin}},\ }\bibfield  {title}
  {\bibinfo {title} {Vortex phase matching as a strategy for schooling in
  robots and in fish},\ }\href@noop {} {\bibfield  {journal} {\bibinfo
  {journal} {Nature Communications}\ }\textbf {\bibinfo {volume} {11}},\
  \bibinfo {pages} {5408} (\bibinfo {year} {2020})}\BibitemShut {NoStop}%
\bibitem [{\citenamefont {Verma}\ \emph {et~al.}(2018)\citenamefont {Verma},
  \citenamefont {Novati},\ and\ \citenamefont
  {Koumoutsakos}}]{verma2018efficient}%
  \BibitemOpen
  \bibfield  {author} {\bibinfo {author} {\bibfnamefont {S.}~\bibnamefont
  {Verma}}, \bibinfo {author} {\bibfnamefont {G.}~\bibnamefont {Novati}},\ and\
  \bibinfo {author} {\bibfnamefont {P.}~\bibnamefont {Koumoutsakos}},\
  }\bibfield  {title} {\bibinfo {title} {Efficient collective swimming by
  harnessing vortices through deep reinforcement learning},\ }\href@noop {}
  {\bibfield  {journal} {\bibinfo  {journal} {Proceedings of the National
  Academy of Sciences}\ }\textbf {\bibinfo {volume} {115}},\ \bibinfo {pages}
  {5849} (\bibinfo {year} {2018})}\BibitemShut {NoStop}%
\bibitem [{\citenamefont {Ashraf}\ \emph {et~al.}(2017)\citenamefont {Ashraf},
  \citenamefont {Bradshaw}, \citenamefont {Ha}, \citenamefont {Halloy},
  \citenamefont {Godoy-Diana},\ and\ \citenamefont
  {Thiria}}]{ashraf2017simple}%
  \BibitemOpen
  \bibfield  {author} {\bibinfo {author} {\bibfnamefont {I.}~\bibnamefont
  {Ashraf}}, \bibinfo {author} {\bibfnamefont {H.}~\bibnamefont {Bradshaw}},
  \bibinfo {author} {\bibfnamefont {T.-T.}\ \bibnamefont {Ha}}, \bibinfo
  {author} {\bibfnamefont {J.}~\bibnamefont {Halloy}}, \bibinfo {author}
  {\bibfnamefont {R.}~\bibnamefont {Godoy-Diana}},\ and\ \bibinfo {author}
  {\bibfnamefont {B.}~\bibnamefont {Thiria}},\ }\bibfield  {title} {\bibinfo
  {title} {Simple phalanx pattern leads to energy saving in cohesive fish
  schooling},\ }\href@noop {} {\bibfield  {journal} {\bibinfo  {journal}
  {Proceedings of the National Academy of Sciences}\ }\textbf {\bibinfo
  {volume} {114}},\ \bibinfo {pages} {9599} (\bibinfo {year}
  {2017})}\BibitemShut {NoStop}%
\bibitem [{\citenamefont {Novati}\ \emph {et~al.}(2017)\citenamefont {Novati},
  \citenamefont {Verma}, \citenamefont {Alexeev}, \citenamefont {Rossinelli},
  \citenamefont {Van~Rees},\ and\ \citenamefont
  {Koumoutsakos}}]{novati2017synchronisation}%
  \BibitemOpen
  \bibfield  {author} {\bibinfo {author} {\bibfnamefont {G.}~\bibnamefont
  {Novati}}, \bibinfo {author} {\bibfnamefont {S.}~\bibnamefont {Verma}},
  \bibinfo {author} {\bibfnamefont {D.}~\bibnamefont {Alexeev}}, \bibinfo
  {author} {\bibfnamefont {D.}~\bibnamefont {Rossinelli}}, \bibinfo {author}
  {\bibfnamefont {W.~M.}\ \bibnamefont {Van~Rees}},\ and\ \bibinfo {author}
  {\bibfnamefont {P.}~\bibnamefont {Koumoutsakos}},\ }\bibfield  {title}
  {\bibinfo {title} {Synchronisation through learning for two self-propelled
  swimmers},\ }\href@noop {} {\bibfield  {journal} {\bibinfo  {journal}
  {Bioinspiration \& Biomimetics}\ }\textbf {\bibinfo {volume} {12}},\ \bibinfo
  {pages} {036001} (\bibinfo {year} {2017})}\BibitemShut {NoStop}%
\bibitem [{\citenamefont {Oza}\ \emph {et~al.}(2019)\citenamefont {Oza},
  \citenamefont {Ristroph},\ and\ \citenamefont {Shelley}}]{oza2019lattices}%
  \BibitemOpen
  \bibfield  {author} {\bibinfo {author} {\bibfnamefont {A.~U.}\ \bibnamefont
  {Oza}}, \bibinfo {author} {\bibfnamefont {L.}~\bibnamefont {Ristroph}},\ and\
  \bibinfo {author} {\bibfnamefont {M.~J.}\ \bibnamefont {Shelley}},\
  }\bibfield  {title} {\bibinfo {title} {Lattices of hydrodynamically
  interacting flapping swimmers},\ }\href@noop {} {\bibfield  {journal}
  {\bibinfo  {journal} {Physical Review X}\ }\textbf {\bibinfo {volume} {9}},\
  \bibinfo {pages} {041024} (\bibinfo {year} {2019})}\BibitemShut {NoStop}%
\bibitem [{\citenamefont {Alben}(2021{\natexlab{a}})}]{alben2021collective1}%
  \BibitemOpen
  \bibfield  {author} {\bibinfo {author} {\bibfnamefont {S.}~\bibnamefont
  {Alben}},\ }\bibfield  {title} {\bibinfo {title} {Collective locomotion of
  two-dimensional lattices of flapping plates. part 1. numerical method,
  single-plate case and lattice input power},\ }\href@noop {} {\bibfield
  {journal} {\bibinfo  {journal} {Journal of Fluid Mechanics}\ }\textbf
  {\bibinfo {volume} {915}},\ \bibinfo {pages} {A20} (\bibinfo {year}
  {2021}{\natexlab{a}})}\BibitemShut {NoStop}%
\bibitem [{\citenamefont {Alben}(2021{\natexlab{b}})}]{alben2021collective2}%
  \BibitemOpen
  \bibfield  {author} {\bibinfo {author} {\bibfnamefont {S.}~\bibnamefont
  {Alben}},\ }\bibfield  {title} {\bibinfo {title} {Collective locomotion of
  two-dimensional lattices of flapping plates. part 2. lattice flows and
  propulsive efficiency},\ }\href@noop {} {\bibfield  {journal} {\bibinfo
  {journal} {Journal of Fluid Mechanics}\ }\textbf {\bibinfo {volume} {915}},\
  \bibinfo {pages} {A21} (\bibinfo {year} {2021}{\natexlab{b}})}\BibitemShut
  {NoStop}%
\bibitem [{\citenamefont {Newbolt}\ \emph {et~al.}(2022)\citenamefont
  {Newbolt}, \citenamefont {Zhang},\ and\ \citenamefont
  {Ristroph}}]{newbolt2022lateral}%
  \BibitemOpen
  \bibfield  {author} {\bibinfo {author} {\bibfnamefont {J.~W.}\ \bibnamefont
  {Newbolt}}, \bibinfo {author} {\bibfnamefont {J.}~\bibnamefont {Zhang}},\
  and\ \bibinfo {author} {\bibfnamefont {L.}~\bibnamefont {Ristroph}},\
  }\bibfield  {title} {\bibinfo {title} {Lateral flow interactions enhance
  speed and stabilize formations of flapping swimmers},\ }\href@noop {}
  {\bibfield  {journal} {\bibinfo  {journal} {Physical Review Fluids}\ }\textbf
  {\bibinfo {volume} {7}},\ \bibinfo {pages} {L061101} (\bibinfo {year}
  {2022})}\BibitemShut {NoStop}%
\bibitem [{\citenamefont {Kurt}\ and\ \citenamefont
  {Moored}(2018)}]{kurt2018flow}%
  \BibitemOpen
  \bibfield  {author} {\bibinfo {author} {\bibfnamefont {M.}~\bibnamefont
  {Kurt}}\ and\ \bibinfo {author} {\bibfnamefont {K.~W.}\ \bibnamefont
  {Moored}},\ }\bibfield  {title} {\bibinfo {title} {Flow interactions of
  two-and three-dimensional networked bio-inspired control elements in an
  in-line arrangement},\ }\href@noop {} {\bibfield  {journal} {\bibinfo
  {journal} {Bioinspiration \& Biomimetics}\ }\textbf {\bibinfo {volume}
  {13}},\ \bibinfo {pages} {045002} (\bibinfo {year} {2018})}\BibitemShut
  {NoStop}%
\bibitem [{\citenamefont {Boschitsch}\ \emph {et~al.}(2014)\citenamefont
  {Boschitsch}, \citenamefont {Dewey},\ and\ \citenamefont
  {Smits}}]{boschitsch2014propulsive}%
  \BibitemOpen
  \bibfield  {author} {\bibinfo {author} {\bibfnamefont {B.~M.}\ \bibnamefont
  {Boschitsch}}, \bibinfo {author} {\bibfnamefont {P.~A.}\ \bibnamefont
  {Dewey}},\ and\ \bibinfo {author} {\bibfnamefont {A.~J.}\ \bibnamefont
  {Smits}},\ }\bibfield  {title} {\bibinfo {title} {Propulsive performance of
  unsteady tandem hydrofoils in an in-line configuration},\ }\href@noop {}
  {\bibfield  {journal} {\bibinfo  {journal} {Physics of Fluids}\ }\textbf
  {\bibinfo {volume} {26}} (\bibinfo {year} {2014})}\BibitemShut {NoStop}%
\bibitem [{\citenamefont {Dewey}\ \emph {et~al.}(2014)\citenamefont {Dewey},
  \citenamefont {Quinn}, \citenamefont {Boschitsch},\ and\ \citenamefont
  {Smits}}]{dewey2014propulsive}%
  \BibitemOpen
  \bibfield  {author} {\bibinfo {author} {\bibfnamefont {P.~A.}\ \bibnamefont
  {Dewey}}, \bibinfo {author} {\bibfnamefont {D.~B.}\ \bibnamefont {Quinn}},
  \bibinfo {author} {\bibfnamefont {B.~M.}\ \bibnamefont {Boschitsch}},\ and\
  \bibinfo {author} {\bibfnamefont {A.~J.}\ \bibnamefont {Smits}},\ }\bibfield
  {title} {\bibinfo {title} {Propulsive performance of unsteady tandem
  hydrofoils in a side-by-side configuration},\ }\href@noop {} {\bibfield
  {journal} {\bibinfo  {journal} {Physics of Fluids}\ }\textbf {\bibinfo
  {volume} {26}} (\bibinfo {year} {2014})}\BibitemShut {NoStop}%
\bibitem [{\citenamefont {Pan}\ and\ \citenamefont
  {Dong}(2020)}]{pan2020computational}%
  \BibitemOpen
  \bibfield  {author} {\bibinfo {author} {\bibfnamefont {Y.}~\bibnamefont
  {Pan}}\ and\ \bibinfo {author} {\bibfnamefont {H.}~\bibnamefont {Dong}},\
  }\bibfield  {title} {\bibinfo {title} {Computational analysis of hydrodynamic
  interactions in a high-density fish school},\ }\href@noop {} {\bibfield
  {journal} {\bibinfo  {journal} {Physics of Fluids}\ }\textbf {\bibinfo
  {volume} {32}} (\bibinfo {year} {2020})}\BibitemShut {NoStop}%
\bibitem [{\citenamefont {Muscutt}\ \emph {et~al.}(2017)\citenamefont
  {Muscutt}, \citenamefont {Weymouth},\ and\ \citenamefont
  {Ganapathisubramani}}]{muscutt2017performance}%
  \BibitemOpen
  \bibfield  {author} {\bibinfo {author} {\bibfnamefont {L.~E.}\ \bibnamefont
  {Muscutt}}, \bibinfo {author} {\bibfnamefont {G.~D.}\ \bibnamefont
  {Weymouth}},\ and\ \bibinfo {author} {\bibfnamefont {B.}~\bibnamefont
  {Ganapathisubramani}},\ }\bibfield  {title} {\bibinfo {title} {Performance
  augmentation mechanism of in-line tandem flapping foils},\ }\href@noop {}
  {\bibfield  {journal} {\bibinfo  {journal} {Journal of Fluid Mechanics}\
  }\textbf {\bibinfo {volume} {827}},\ \bibinfo {pages} {484} (\bibinfo {year}
  {2017})}\BibitemShut {NoStop}%
\bibitem [{\citenamefont {Seo}\ and\ \citenamefont
  {Mittal}(2022)}]{seo2022improved}%
  \BibitemOpen
  \bibfield  {author} {\bibinfo {author} {\bibfnamefont {J.-H.}\ \bibnamefont
  {Seo}}\ and\ \bibinfo {author} {\bibfnamefont {R.}~\bibnamefont {Mittal}},\
  }\bibfield  {title} {\bibinfo {title} {Improved swimming performance in
  schooling fish via leading-edge vortex enhancement},\ }\href@noop {}
  {\bibfield  {journal} {\bibinfo  {journal} {Bioinspiration \& Biomimetics}\
  }\textbf {\bibinfo {volume} {17}},\ \bibinfo {pages} {066020} (\bibinfo
  {year} {2022})}\BibitemShut {NoStop}%
\bibitem [{\citenamefont {Pan}\ and\ \citenamefont
  {Dong}(2022)}]{pan2022effects}%
  \BibitemOpen
  \bibfield  {author} {\bibinfo {author} {\bibfnamefont {Y.}~\bibnamefont
  {Pan}}\ and\ \bibinfo {author} {\bibfnamefont {H.}~\bibnamefont {Dong}},\
  }\bibfield  {title} {\bibinfo {title} {Effects of phase difference on
  hydrodynamic interactions and wake patterns in high-density fish schools},\
  }\href@noop {} {\bibfield  {journal} {\bibinfo  {journal} {Physics of
  Fluids}\ }\textbf {\bibinfo {volume} {34}} (\bibinfo {year}
  {2022})}\BibitemShut {NoStop}%
\bibitem [{\citenamefont {Saadat}\ \emph {et~al.}(2021)\citenamefont {Saadat},
  \citenamefont {Berlinger}, \citenamefont {Sheshmani}, \citenamefont {Nagpal},
  \citenamefont {Lauder},\ and\ \citenamefont
  {Haj-Hariri}}]{saadat2021hydrodynamic}%
  \BibitemOpen
  \bibfield  {author} {\bibinfo {author} {\bibfnamefont {M.}~\bibnamefont
  {Saadat}}, \bibinfo {author} {\bibfnamefont {F.}~\bibnamefont {Berlinger}},
  \bibinfo {author} {\bibfnamefont {A.}~\bibnamefont {Sheshmani}}, \bibinfo
  {author} {\bibfnamefont {R.}~\bibnamefont {Nagpal}}, \bibinfo {author}
  {\bibfnamefont {G.~V.}\ \bibnamefont {Lauder}},\ and\ \bibinfo {author}
  {\bibfnamefont {H.}~\bibnamefont {Haj-Hariri}},\ }\bibfield  {title}
  {\bibinfo {title} {Hydrodynamic advantages of in-line schooling},\
  }\href@noop {} {\bibfield  {journal} {\bibinfo  {journal} {Bioinspiration \&
  Biomimetics}\ }\textbf {\bibinfo {volume} {16}},\ \bibinfo {pages} {046002}
  (\bibinfo {year} {2021})}\BibitemShut {NoStop}%
\bibitem [{\citenamefont {Couzin}\ and\ \citenamefont
  {Krause}(2003)}]{couzin2003self}%
  \BibitemOpen
  \bibfield  {author} {\bibinfo {author} {\bibfnamefont {I.~D.}\ \bibnamefont
  {Couzin}}\ and\ \bibinfo {author} {\bibfnamefont {J.}~\bibnamefont
  {Krause}},\ }\bibfield  {title} {\bibinfo {title} {Self-organization and
  collective behavior in vertebrates},\ }\href@noop {} {\bibfield  {journal}
  {\bibinfo  {journal} {Advances in the Study of Behavior}\ }\textbf {\bibinfo
  {volume} {32}},\ \bibinfo {pages} {10} (\bibinfo {year} {2003})}\BibitemShut
  {NoStop}%
\bibitem [{\citenamefont {Couzin}\ \emph {et~al.}(2005)\citenamefont {Couzin},
  \citenamefont {Krause}, \citenamefont {Franks},\ and\ \citenamefont
  {Levin}}]{couzin2005effective}%
  \BibitemOpen
  \bibfield  {author} {\bibinfo {author} {\bibfnamefont {I.~D.}\ \bibnamefont
  {Couzin}}, \bibinfo {author} {\bibfnamefont {J.}~\bibnamefont {Krause}},
  \bibinfo {author} {\bibfnamefont {N.~R.}\ \bibnamefont {Franks}},\ and\
  \bibinfo {author} {\bibfnamefont {S.~A.}\ \bibnamefont {Levin}},\ }\bibfield
  {title} {\bibinfo {title} {Effective leadership and decision-making in animal
  groups on the move},\ }\href@noop {} {\bibfield  {journal} {\bibinfo
  {journal} {Nature}\ }\textbf {\bibinfo {volume} {433}},\ \bibinfo {pages}
  {513} (\bibinfo {year} {2005})}\BibitemShut {NoStop}%
\bibitem [{\citenamefont {Breder}(1954)}]{breder1954equations}%
  \BibitemOpen
  \bibfield  {author} {\bibinfo {author} {\bibfnamefont {C.~M.}\ \bibnamefont
  {Breder}},\ }\bibfield  {title} {\bibinfo {title} {Equations descriptive of
  fish schools and other animal aggregations},\ }\href@noop {} {\bibfield
  {journal} {\bibinfo  {journal} {Ecology}\ }\textbf {\bibinfo {volume} {35}},\
  \bibinfo {pages} {361} (\bibinfo {year} {1954})}\BibitemShut {NoStop}%
\bibitem [{\citenamefont {Aoki}(1982)}]{19821081}%
  \BibitemOpen
  \bibfield  {author} {\bibinfo {author} {\bibfnamefont {I.}~\bibnamefont
  {Aoki}},\ }\bibfield  {title} {\bibinfo {title} {A simulation study on the
  schooling mechanism in fish},\ }\href
  {https://doi.org/10.2331/suisan.48.1081} {\bibfield  {journal} {\bibinfo
  {journal} {NIPPON SUISAN GAKKAISHI}\ }\textbf {\bibinfo {volume} {48}},\
  \bibinfo {pages} {1081} (\bibinfo {year} {1982})}\BibitemShut {NoStop}%
\bibitem [{\citenamefont {Reynolds}(1987)}]{reynolds1987flocks}%
  \BibitemOpen
  \bibfield  {author} {\bibinfo {author} {\bibfnamefont {C.~W.}\ \bibnamefont
  {Reynolds}},\ }\bibfield  {title} {\bibinfo {title} {Flocks, herds and
  schools: A distributed behavioral model},\ }in\ \href@noop {} {\emph
  {\bibinfo {booktitle} {Proceedings of the 14th annual conference on Computer
  graphics and interactive techniques}}}\ (\bibinfo {year} {1987})\ pp.\
  \bibinfo {pages} {25--34}\BibitemShut {NoStop}%
\bibitem [{\citenamefont {Couzin}\ \emph {et~al.}(2002)\citenamefont {Couzin},
  \citenamefont {Krause}, \citenamefont {James}, \citenamefont {Ruxton},\ and\
  \citenamefont {Franks}}]{couzin2002collective}%
  \BibitemOpen
  \bibfield  {author} {\bibinfo {author} {\bibfnamefont {I.~D.}\ \bibnamefont
  {Couzin}}, \bibinfo {author} {\bibfnamefont {J.}~\bibnamefont {Krause}},
  \bibinfo {author} {\bibfnamefont {R.}~\bibnamefont {James}}, \bibinfo
  {author} {\bibfnamefont {G.~D.}\ \bibnamefont {Ruxton}},\ and\ \bibinfo
  {author} {\bibfnamefont {N.~R.}\ \bibnamefont {Franks}},\ }\bibfield  {title}
  {\bibinfo {title} {Collective memory and spatial sorting in animal groups},\
  }\href@noop {} {\bibfield  {journal} {\bibinfo  {journal} {Journal of
  Theoretical Biology}\ }\textbf {\bibinfo {volume} {218}},\ \bibinfo {pages}
  {1} (\bibinfo {year} {2002})}\BibitemShut {NoStop}%
\bibitem [{\citenamefont {Ko}\ \emph {et~al.}(2023)\citenamefont {Ko},
  \citenamefont {Lauder},\ and\ \citenamefont {Nagpal}}]{ko2023role}%
  \BibitemOpen
  \bibfield  {author} {\bibinfo {author} {\bibfnamefont {H.}~\bibnamefont
  {Ko}}, \bibinfo {author} {\bibfnamefont {G.}~\bibnamefont {Lauder}},\ and\
  \bibinfo {author} {\bibfnamefont {R.}~\bibnamefont {Nagpal}},\ }\bibfield
  {title} {\bibinfo {title} {The role of hydrodynamics in collective motions of
  fish schools and bioinspired underwater robots},\ }\href@noop {} {\bibfield
  {journal} {\bibinfo  {journal} {Journal of the Royal Society Interface}\
  }\textbf {\bibinfo {volume} {20}},\ \bibinfo {pages} {20230357} (\bibinfo
  {year} {2023})}\BibitemShut {NoStop}%
\bibitem [{\citenamefont {Filella}\ \emph {et~al.}(2018)\citenamefont
  {Filella}, \citenamefont {Nadal}, \citenamefont {Sire}, \citenamefont
  {Kanso},\ and\ \citenamefont {Eloy}}]{filella2018model}%
  \BibitemOpen
  \bibfield  {author} {\bibinfo {author} {\bibfnamefont {A.}~\bibnamefont
  {Filella}}, \bibinfo {author} {\bibfnamefont {F.}~\bibnamefont {Nadal}},
  \bibinfo {author} {\bibfnamefont {C.}~\bibnamefont {Sire}}, \bibinfo {author}
  {\bibfnamefont {E.}~\bibnamefont {Kanso}},\ and\ \bibinfo {author}
  {\bibfnamefont {C.}~\bibnamefont {Eloy}},\ }\bibfield  {title} {\bibinfo
  {title} {Model of collective fish behavior with hydrodynamic interactions},\
  }\href@noop {} {\bibfield  {journal} {\bibinfo  {journal} {Physical Review
  Letters}\ }\textbf {\bibinfo {volume} {120}},\ \bibinfo {pages} {198101}
  (\bibinfo {year} {2018})}\BibitemShut {NoStop}%
\bibitem [{\citenamefont {Lighthill}(1975)}]{lighthill1975mathematical}%
  \BibitemOpen
  \bibfield  {author} {\bibinfo {author} {\bibfnamefont {J.}~\bibnamefont
  {Lighthill}},\ }\href@noop {} {\emph {\bibinfo {title} {Mathematical
  biofluiddynamics}}}\ (\bibinfo  {publisher} {SIAM},\ \bibinfo {year}
  {1975})\BibitemShut {NoStop}%
\bibitem [{\citenamefont {Kelly}(1959)}]{kelly1959two}%
  \BibitemOpen
  \bibfield  {author} {\bibinfo {author} {\bibfnamefont {H.~R.}\ \bibnamefont
  {Kelly}},\ }\href@noop {} {\emph {\bibinfo {title} {A two-body problem in the
  echelon-formation swimming of porpoise}}},\ \bibinfo {type} {Tech. Rep.}\
  \bibinfo {number} {40606-1}\ (\bibinfo  {institution} {U.S. Naval Ordinance
  Testing Station},\ \bibinfo {address} {China Lake, CA},\ \bibinfo {year}
  {1959})\BibitemShut {NoStop}%
\bibitem [{\citenamefont {Tchieu}\ \emph {et~al.}(2012)\citenamefont {Tchieu},
  \citenamefont {Kanso},\ and\ \citenamefont {Newton}}]{tchieu2012finite}%
  \BibitemOpen
  \bibfield  {author} {\bibinfo {author} {\bibfnamefont {A.~A.}\ \bibnamefont
  {Tchieu}}, \bibinfo {author} {\bibfnamefont {E.}~\bibnamefont {Kanso}},\ and\
  \bibinfo {author} {\bibfnamefont {P.~K.}\ \bibnamefont {Newton}},\ }\bibfield
   {title} {\bibinfo {title} {The finite-dipole dynamical system},\ }\href@noop
  {} {\bibfield  {journal} {\bibinfo  {journal} {Proceedings of the Royal
  Society A: Mathematical, Physical and Engineering Sciences}\ }\textbf
  {\bibinfo {volume} {468}},\ \bibinfo {pages} {3006} (\bibinfo {year}
  {2012})}\BibitemShut {NoStop}%
\bibitem [{\citenamefont {Kanso}\ and\ \citenamefont
  {Tsang}(2014)}]{kanso2014dipole}%
  \BibitemOpen
  \bibfield  {author} {\bibinfo {author} {\bibfnamefont {E.}~\bibnamefont
  {Kanso}}\ and\ \bibinfo {author} {\bibfnamefont {A.~C.~H.}\ \bibnamefont
  {Tsang}},\ }\bibfield  {title} {\bibinfo {title} {Dipole models of
  self-propelled bodies},\ }\href@noop {} {\bibfield  {journal} {\bibinfo
  {journal} {Fluid Dynamics Research}\ }\textbf {\bibinfo {volume} {46}},\
  \bibinfo {pages} {061407} (\bibinfo {year} {2014})}\BibitemShut {NoStop}%
\bibitem [{\citenamefont {Gazzola}\ \emph {et~al.}(2016)\citenamefont
  {Gazzola}, \citenamefont {Tchieu}, \citenamefont {Alexeev}, \citenamefont
  {De~Brauer},\ and\ \citenamefont {Koumoutsakos}}]{gazzola2016learning}%
  \BibitemOpen
  \bibfield  {author} {\bibinfo {author} {\bibfnamefont {M.}~\bibnamefont
  {Gazzola}}, \bibinfo {author} {\bibfnamefont {A.~A.}\ \bibnamefont {Tchieu}},
  \bibinfo {author} {\bibfnamefont {D.}~\bibnamefont {Alexeev}}, \bibinfo
  {author} {\bibfnamefont {A.}~\bibnamefont {De~Brauer}},\ and\ \bibinfo
  {author} {\bibfnamefont {P.}~\bibnamefont {Koumoutsakos}},\ }\bibfield
  {title} {\bibinfo {title} {Learning to school in the presence of hydrodynamic
  interactions},\ }\href@noop {} {\bibfield  {journal} {\bibinfo  {journal}
  {Journal of Fluid Mechanics}\ }\textbf {\bibinfo {volume} {789}},\ \bibinfo
  {pages} {726} (\bibinfo {year} {2016})}\BibitemShut {NoStop}%
\bibitem [{\citenamefont {Porfiri}\ \emph {et~al.}(2022)\citenamefont
  {Porfiri}, \citenamefont {Zhang},\ and\ \citenamefont
  {Peterson}}]{porfiri2022hydrodynamic}%
  \BibitemOpen
  \bibfield  {author} {\bibinfo {author} {\bibfnamefont {M.}~\bibnamefont
  {Porfiri}}, \bibinfo {author} {\bibfnamefont {P.}~\bibnamefont {Zhang}},\
  and\ \bibinfo {author} {\bibfnamefont {S.~D.}\ \bibnamefont {Peterson}},\
  }\bibfield  {title} {\bibinfo {title} {Hydrodynamic model of fish orientation
  in a channel flow},\ }\href@noop {} {\bibfield  {journal} {\bibinfo
  {journal} {eLife}\ }\textbf {\bibinfo {volume} {11}},\ \bibinfo {pages}
  {e75225} (\bibinfo {year} {2022})}\BibitemShut {NoStop}%
\bibitem [{\citenamefont {Becker}\ \emph {et~al.}(2015)\citenamefont {Becker},
  \citenamefont {Masoud}, \citenamefont {Newbolt}, \citenamefont {Shelley},\
  and\ \citenamefont {Ristroph}}]{becker2015hydrodynamic}%
  \BibitemOpen
  \bibfield  {author} {\bibinfo {author} {\bibfnamefont {A.~D.}\ \bibnamefont
  {Becker}}, \bibinfo {author} {\bibfnamefont {H.}~\bibnamefont {Masoud}},
  \bibinfo {author} {\bibfnamefont {J.~W.}\ \bibnamefont {Newbolt}}, \bibinfo
  {author} {\bibfnamefont {M.}~\bibnamefont {Shelley}},\ and\ \bibinfo {author}
  {\bibfnamefont {L.}~\bibnamefont {Ristroph}},\ }\bibfield  {title} {\bibinfo
  {title} {Hydrodynamic schooling of flapping swimmers},\ }\href@noop {}
  {\bibfield  {journal} {\bibinfo  {journal} {Nature Communications}\ }\textbf
  {\bibinfo {volume} {6}},\ \bibinfo {pages} {8514} (\bibinfo {year}
  {2015})}\BibitemShut {NoStop}%
\bibitem [{\citenamefont {Ramananarivo}\ \emph {et~al.}(2016)\citenamefont
  {Ramananarivo}, \citenamefont {Fang}, \citenamefont {Oza}, \citenamefont
  {Zhang},\ and\ \citenamefont {Ristroph}}]{ramananarivo2016flow}%
  \BibitemOpen
  \bibfield  {author} {\bibinfo {author} {\bibfnamefont {S.}~\bibnamefont
  {Ramananarivo}}, \bibinfo {author} {\bibfnamefont {F.}~\bibnamefont {Fang}},
  \bibinfo {author} {\bibfnamefont {A.}~\bibnamefont {Oza}}, \bibinfo {author}
  {\bibfnamefont {J.}~\bibnamefont {Zhang}},\ and\ \bibinfo {author}
  {\bibfnamefont {L.}~\bibnamefont {Ristroph}},\ }\bibfield  {title} {\bibinfo
  {title} {Flow interactions lead to orderly formations of flapping wings in
  forward flight},\ }\href@noop {} {\bibfield  {journal} {\bibinfo  {journal}
  {Physical Review Fluids}\ }\textbf {\bibinfo {volume} {1}},\ \bibinfo {pages}
  {071201} (\bibinfo {year} {2016})}\BibitemShut {NoStop}%
\bibitem [{\citenamefont {Newbolt}\ \emph {et~al.}(2019)\citenamefont
  {Newbolt}, \citenamefont {Zhang},\ and\ \citenamefont
  {Ristroph}}]{newbolt2019flow}%
  \BibitemOpen
  \bibfield  {author} {\bibinfo {author} {\bibfnamefont {J.~W.}\ \bibnamefont
  {Newbolt}}, \bibinfo {author} {\bibfnamefont {J.}~\bibnamefont {Zhang}},\
  and\ \bibinfo {author} {\bibfnamefont {L.}~\bibnamefont {Ristroph}},\
  }\bibfield  {title} {\bibinfo {title} {Flow interactions between
  uncoordinated flapping swimmers give rise to group cohesion},\ }\href@noop {}
  {\bibfield  {journal} {\bibinfo  {journal} {Proceedings of the National
  Academy of Sciences}\ }\textbf {\bibinfo {volume} {116}},\ \bibinfo {pages}
  {2419} (\bibinfo {year} {2019})}\BibitemShut {NoStop}%
\bibitem [{\citenamefont {Gazzola}\ \emph
  {et~al.}(2014{\natexlab{a}})\citenamefont {Gazzola}, \citenamefont
  {Hejazialhosseini},\ and\ \citenamefont
  {Koumoutsakos}}]{gazzola2014reinforcement}%
  \BibitemOpen
  \bibfield  {author} {\bibinfo {author} {\bibfnamefont {M.}~\bibnamefont
  {Gazzola}}, \bibinfo {author} {\bibfnamefont {B.}~\bibnamefont
  {Hejazialhosseini}},\ and\ \bibinfo {author} {\bibfnamefont {P.}~\bibnamefont
  {Koumoutsakos}},\ }\bibfield  {title} {\bibinfo {title} {Reinforcement
  learning and wavelet adapted vortex methods for simulations of self-propelled
  swimmers},\ }\href@noop {} {\bibfield  {journal} {\bibinfo  {journal} {SIAM
  Journal on Scientific Computing}\ }\textbf {\bibinfo {volume} {36}},\
  \bibinfo {pages} {B622} (\bibinfo {year} {2014}{\natexlab{a}})}\BibitemShut
  {NoStop}%
\bibitem [{\citenamefont {Lin}\ \emph {et~al.}(2021)\citenamefont {Lin},
  \citenamefont {Wu}, \citenamefont {Zhang},\ and\ \citenamefont
  {Yang}}]{lin2021flow}%
  \BibitemOpen
  \bibfield  {author} {\bibinfo {author} {\bibfnamefont {X.}~\bibnamefont
  {Lin}}, \bibinfo {author} {\bibfnamefont {J.}~\bibnamefont {Wu}}, \bibinfo
  {author} {\bibfnamefont {T.}~\bibnamefont {Zhang}},\ and\ \bibinfo {author}
  {\bibfnamefont {L.}~\bibnamefont {Yang}},\ }\bibfield  {title} {\bibinfo
  {title} {Flow-mediated organization of two freely flapping swimmers},\
  }\href@noop {} {\bibfield  {journal} {\bibinfo  {journal} {Journal of Fluid
  Mechanics}\ }\textbf {\bibinfo {volume} {912}},\ \bibinfo {pages} {A37}
  (\bibinfo {year} {2021})}\BibitemShut {NoStop}%
\bibitem [{\citenamefont {Lin}\ \emph {et~al.}(2022)\citenamefont {Lin},
  \citenamefont {Wu}, \citenamefont {Yang},\ and\ \citenamefont
  {Dong}}]{lin2022two}%
  \BibitemOpen
  \bibfield  {author} {\bibinfo {author} {\bibfnamefont {X.}~\bibnamefont
  {Lin}}, \bibinfo {author} {\bibfnamefont {J.}~\bibnamefont {Wu}}, \bibinfo
  {author} {\bibfnamefont {L.}~\bibnamefont {Yang}},\ and\ \bibinfo {author}
  {\bibfnamefont {H.}~\bibnamefont {Dong}},\ }\bibfield  {title} {\bibinfo
  {title} {Two-dimensional hydrodynamic schooling of two flapping swimmers
  initially in tandem formation},\ }\href@noop {} {\bibfield  {journal}
  {\bibinfo  {journal} {Journal of Fluid Mechanics}\ }\textbf {\bibinfo
  {volume} {941}},\ \bibinfo {pages} {A29} (\bibinfo {year}
  {2022})}\BibitemShut {NoStop}%
\bibitem [{\citenamefont {Ormonde}\ \emph {et~al.}(2021)\citenamefont
  {Ormonde}, \citenamefont {Kurt}, \citenamefont {Mivehchi},\ and\
  \citenamefont {Moored}}]{ormonde2021two}%
  \BibitemOpen
  \bibfield  {author} {\bibinfo {author} {\bibfnamefont {P.~C.}\ \bibnamefont
  {Ormonde}}, \bibinfo {author} {\bibfnamefont {M.}~\bibnamefont {Kurt}},
  \bibinfo {author} {\bibfnamefont {A.}~\bibnamefont {Mivehchi}},\ and\
  \bibinfo {author} {\bibfnamefont {K.~W.}\ \bibnamefont {Moored}},\ }\bibfield
   {title} {\bibinfo {title} {Two-dimensionally stable self-organization arises
  in simple schooling swimmers through hydrodynamic interactions},\ }\href@noop
  {} {\bibfield  {journal} {\bibinfo  {journal} {arXiv preprint
  arXiv:2102.03571}\ } (\bibinfo {year} {2021})}\BibitemShut {NoStop}%
\bibitem [{\citenamefont {Eldredge}(2019)}]{eldredge2019mathematical}%
  \BibitemOpen
  \bibfield  {author} {\bibinfo {author} {\bibfnamefont {J.~D.}\ \bibnamefont
  {Eldredge}},\ }\href@noop {} {\emph {\bibinfo {title} {Mathematical Modeling
  of Unsteady Inviscid Flows}}},\ Vol.~\bibinfo {volume} {50}\ (\bibinfo
  {publisher} {Springer},\ \bibinfo {year} {2019})\BibitemShut {NoStop}%
\bibitem [{\citenamefont {Batchelor}(2000)}]{batchelor2000introduction}%
  \BibitemOpen
  \bibfield  {author} {\bibinfo {author} {\bibfnamefont {G.~K.}\ \bibnamefont
  {Batchelor}},\ }\href@noop {} {\emph {\bibinfo {title} {An Introduction to
  Fluid Dynamics}}}\ (\bibinfo  {publisher} {Cambridge University Press},\
  \bibinfo {year} {2000})\BibitemShut {NoStop}%
\bibitem [{\citenamefont {Zhu}\ \emph {et~al.}(2002)\citenamefont {Zhu},
  \citenamefont {Wolfgang}, \citenamefont {Yue},\ and\ \citenamefont
  {Triantafyllou}}]{zhu2002three}%
  \BibitemOpen
  \bibfield  {author} {\bibinfo {author} {\bibfnamefont {Q.}~\bibnamefont
  {Zhu}}, \bibinfo {author} {\bibfnamefont {M.}~\bibnamefont {Wolfgang}},
  \bibinfo {author} {\bibfnamefont {D.}~\bibnamefont {Yue}},\ and\ \bibinfo
  {author} {\bibfnamefont {M.}~\bibnamefont {Triantafyllou}},\ }\bibfield
  {title} {\bibinfo {title} {Three-dimensional flow structures and vorticity
  control in fish-like swimming},\ }\href@noop {} {\bibfield  {journal}
  {\bibinfo  {journal} {Journal of Fluid Mechanics}\ }\textbf {\bibinfo
  {volume} {468}},\ \bibinfo {pages} {1} (\bibinfo {year} {2002})}\BibitemShut
  {NoStop}%
\bibitem [{\citenamefont {Kimura}\ and\ \citenamefont
  {Moffatt}(2017)}]{kimura2017scaling}%
  \BibitemOpen
  \bibfield  {author} {\bibinfo {author} {\bibfnamefont {Y.}~\bibnamefont
  {Kimura}}\ and\ \bibinfo {author} {\bibfnamefont {H.}~\bibnamefont
  {Moffatt}},\ }\bibfield  {title} {\bibinfo {title} {Scaling properties
  towards vortex reconnection under {Bi}ot--{S}avart evolution},\ }\href@noop
  {} {\bibfield  {journal} {\bibinfo  {journal} {Fluid Dynamics Research}\
  }\textbf {\bibinfo {volume} {50}},\ \bibinfo {pages} {011409} (\bibinfo
  {year} {2017})}\BibitemShut {NoStop}%
\bibitem [{\citenamefont {Saffman}(1993)}]{saffman_1993}%
  \BibitemOpen
  \bibfield  {author} {\bibinfo {author} {\bibfnamefont {P.~G.}\ \bibnamefont
  {Saffman}},\ }\href {https://doi.org/10.1017/CBO9780511624063} {\emph
  {\bibinfo {title} {Vortex Dynamics}}},\ Cambridge Monographs on Mechanics\
  (\bibinfo  {publisher} {Cambridge University Press},\ \bibinfo {year}
  {1993})\BibitemShut {NoStop}%
\bibitem [{\citenamefont {Hernandez-Ortiz}\ \emph {et~al.}(2005)\citenamefont
  {Hernandez-Ortiz}, \citenamefont {Stoltz},\ and\ \citenamefont
  {Graham}}]{hernandez2005transport}%
  \BibitemOpen
  \bibfield  {author} {\bibinfo {author} {\bibfnamefont {J.~P.}\ \bibnamefont
  {Hernandez-Ortiz}}, \bibinfo {author} {\bibfnamefont {C.~G.}\ \bibnamefont
  {Stoltz}},\ and\ \bibinfo {author} {\bibfnamefont {M.~D.}\ \bibnamefont
  {Graham}},\ }\bibfield  {title} {\bibinfo {title} {Transport and collective
  dynamics in suspensions of confined swimming particles},\ }\href@noop {}
  {\bibfield  {journal} {\bibinfo  {journal} {Physical Review Letters}\
  }\textbf {\bibinfo {volume} {95}},\ \bibinfo {pages} {204501} (\bibinfo
  {year} {2005})}\BibitemShut {NoStop}%
\bibitem [{\citenamefont {Hern{\'a}ndez-Ortiz}\ \emph
  {et~al.}(2007)\citenamefont {Hern{\'a}ndez-Ortiz}, \citenamefont {de~Pablo},\
  and\ \citenamefont {Graham}}]{hernandez2007fast}%
  \BibitemOpen
  \bibfield  {author} {\bibinfo {author} {\bibfnamefont {J.~P.}\ \bibnamefont
  {Hern{\'a}ndez-Ortiz}}, \bibinfo {author} {\bibfnamefont {J.~J.}\
  \bibnamefont {de~Pablo}},\ and\ \bibinfo {author} {\bibfnamefont {M.~D.}\
  \bibnamefont {Graham}},\ }\bibfield  {title} {\bibinfo {title} {Fast
  computation of many-particle hydrodynamic and electrostatic interactions in a
  confined geometry},\ }\href@noop {} {\bibfield  {journal} {\bibinfo
  {journal} {Physical Review Letters}\ }\textbf {\bibinfo {volume} {98}},\
  \bibinfo {pages} {140602} (\bibinfo {year} {2007})}\BibitemShut {NoStop}%
\bibitem [{\citenamefont {Underhill}\ \emph {et~al.}(2008)\citenamefont
  {Underhill}, \citenamefont {Hernandez-Ortiz},\ and\ \citenamefont
  {Graham}}]{underhill2008diffusion}%
  \BibitemOpen
  \bibfield  {author} {\bibinfo {author} {\bibfnamefont {P.~T.}\ \bibnamefont
  {Underhill}}, \bibinfo {author} {\bibfnamefont {J.~P.}\ \bibnamefont
  {Hernandez-Ortiz}},\ and\ \bibinfo {author} {\bibfnamefont {M.~D.}\
  \bibnamefont {Graham}},\ }\bibfield  {title} {\bibinfo {title} {Diffusion and
  spatial correlations in suspensions of swimming particles},\ }\href@noop {}
  {\bibfield  {journal} {\bibinfo  {journal} {Physical Review Letters}\
  }\textbf {\bibinfo {volume} {100}},\ \bibinfo {pages} {248101} (\bibinfo
  {year} {2008})}\BibitemShut {NoStop}%
\bibitem [{\citenamefont {Hernandez-Ortiz}\ \emph {et~al.}(2009)\citenamefont
  {Hernandez-Ortiz}, \citenamefont {Underhill},\ and\ \citenamefont
  {Graham}}]{hernandez2009dynamics}%
  \BibitemOpen
  \bibfield  {author} {\bibinfo {author} {\bibfnamefont {J.~P.}\ \bibnamefont
  {Hernandez-Ortiz}}, \bibinfo {author} {\bibfnamefont {P.~T.}\ \bibnamefont
  {Underhill}},\ and\ \bibinfo {author} {\bibfnamefont {M.~D.}\ \bibnamefont
  {Graham}},\ }\bibfield  {title} {\bibinfo {title} {Dynamics of confined
  suspensions of swimming particles},\ }\href@noop {} {\bibfield  {journal}
  {\bibinfo  {journal} {Journal of Physics: Condensed Matter}\ }\textbf
  {\bibinfo {volume} {21}},\ \bibinfo {pages} {204107} (\bibinfo {year}
  {2009})}\BibitemShut {NoStop}%
\bibitem [{\citenamefont {Ryan}\ \emph {et~al.}(2011)\citenamefont {Ryan},
  \citenamefont {Haines}, \citenamefont {Berlyand}, \citenamefont {Ziebert},\
  and\ \citenamefont {Aranson}}]{ryan2011viscosity}%
  \BibitemOpen
  \bibfield  {author} {\bibinfo {author} {\bibfnamefont {S.~D.}\ \bibnamefont
  {Ryan}}, \bibinfo {author} {\bibfnamefont {B.~M.}\ \bibnamefont {Haines}},
  \bibinfo {author} {\bibfnamefont {L.}~\bibnamefont {Berlyand}}, \bibinfo
  {author} {\bibfnamefont {F.}~\bibnamefont {Ziebert}},\ and\ \bibinfo {author}
  {\bibfnamefont {I.~S.}\ \bibnamefont {Aranson}},\ }\bibfield  {title}
  {\bibinfo {title} {Viscosity of bacterial suspensions: Hydrodynamic
  interactions and self-induced noise},\ }\href@noop {} {\bibfield  {journal}
  {\bibinfo  {journal} {Physical Review E}\ }\textbf {\bibinfo {volume} {83}},\
  \bibinfo {pages} {050904} (\bibinfo {year} {2011})}\BibitemShut {NoStop}%
\bibitem [{\citenamefont {Auton}\ \emph {et~al.}(1988)\citenamefont {Auton},
  \citenamefont {Hunt},\ and\ \citenamefont {Prud'Homme}}]{auton1988force}%
  \BibitemOpen
  \bibfield  {author} {\bibinfo {author} {\bibfnamefont {T.~R.}\ \bibnamefont
  {Auton}}, \bibinfo {author} {\bibfnamefont {J.~C.~R.}\ \bibnamefont {Hunt}},\
  and\ \bibinfo {author} {\bibfnamefont {M.}~\bibnamefont {Prud'Homme}},\
  }\bibfield  {title} {\bibinfo {title} {The force exerted on a body in
  inviscid unsteady non-uniform rotational flow},\ }\href@noop {} {\bibfield
  {journal} {\bibinfo  {journal} {Journal of Fluid Mechanics}\ }\textbf
  {\bibinfo {volume} {197}},\ \bibinfo {pages} {241} (\bibinfo {year}
  {1988})}\BibitemShut {NoStop}%
\bibitem [{\citenamefont {Huang}\ and\ \citenamefont
  {Fauci}(2017)}]{huang2017interaction}%
  \BibitemOpen
  \bibfield  {author} {\bibinfo {author} {\bibfnamefont {J.}~\bibnamefont
  {Huang}}\ and\ \bibinfo {author} {\bibfnamefont {L.}~\bibnamefont {Fauci}},\
  }\bibfield  {title} {\bibinfo {title} {Interaction of toroidal swimmers in
  {S}tokes flow},\ }\href@noop {} {\bibfield  {journal} {\bibinfo  {journal}
  {Physical Review E}\ }\textbf {\bibinfo {volume} {95}},\ \bibinfo {pages}
  {043102} (\bibinfo {year} {2017})}\BibitemShut {NoStop}%
\bibitem [{\citenamefont {Chatzimanolakis}(2023)}]{chatzimanolakis2023high}%
  \BibitemOpen
  \bibfield  {author} {\bibinfo {author} {\bibfnamefont {M.}~\bibnamefont
  {Chatzimanolakis}},\ }\emph {\bibinfo {title} {High Performance Computing of
  Vortex Dominated Flows with Adaptive Mesh Refinement}},\ \href@noop {} {Ph.D.
  thesis},\ \bibinfo  {school} {ETH Zurich} (\bibinfo {year}
  {2023})\BibitemShut {NoStop}%
\bibitem [{\citenamefont {Gazzola}\ \emph
  {et~al.}(2014{\natexlab{b}})\citenamefont {Gazzola}, \citenamefont
  {Argentina},\ and\ \citenamefont {Mahadevan}}]{gazzola2014scaling}%
  \BibitemOpen
  \bibfield  {author} {\bibinfo {author} {\bibfnamefont {M.}~\bibnamefont
  {Gazzola}}, \bibinfo {author} {\bibfnamefont {M.}~\bibnamefont {Argentina}},\
  and\ \bibinfo {author} {\bibfnamefont {L.}~\bibnamefont {Mahadevan}},\
  }\bibfield  {title} {\bibinfo {title} {Scaling macroscopic aquatic
  locomotion},\ }\href@noop {} {\bibfield  {journal} {\bibinfo  {journal}
  {Nature Physics}\ }\textbf {\bibinfo {volume} {10}},\ \bibinfo {pages} {758}
  (\bibinfo {year} {2014}{\natexlab{b}})}\BibitemShut {NoStop}%
\end{thebibliography}%

\newpage
\appendix
\renewcommand\thefigure{\thesection\arabic{figure}}
\counterwithin{figure}{section}
\section{\label{sec5:AppPD3D}Phase diagrams for three-dimensional model}
\begin{figure}[!h]
    \centering
    \includegraphics[scale=0.64]{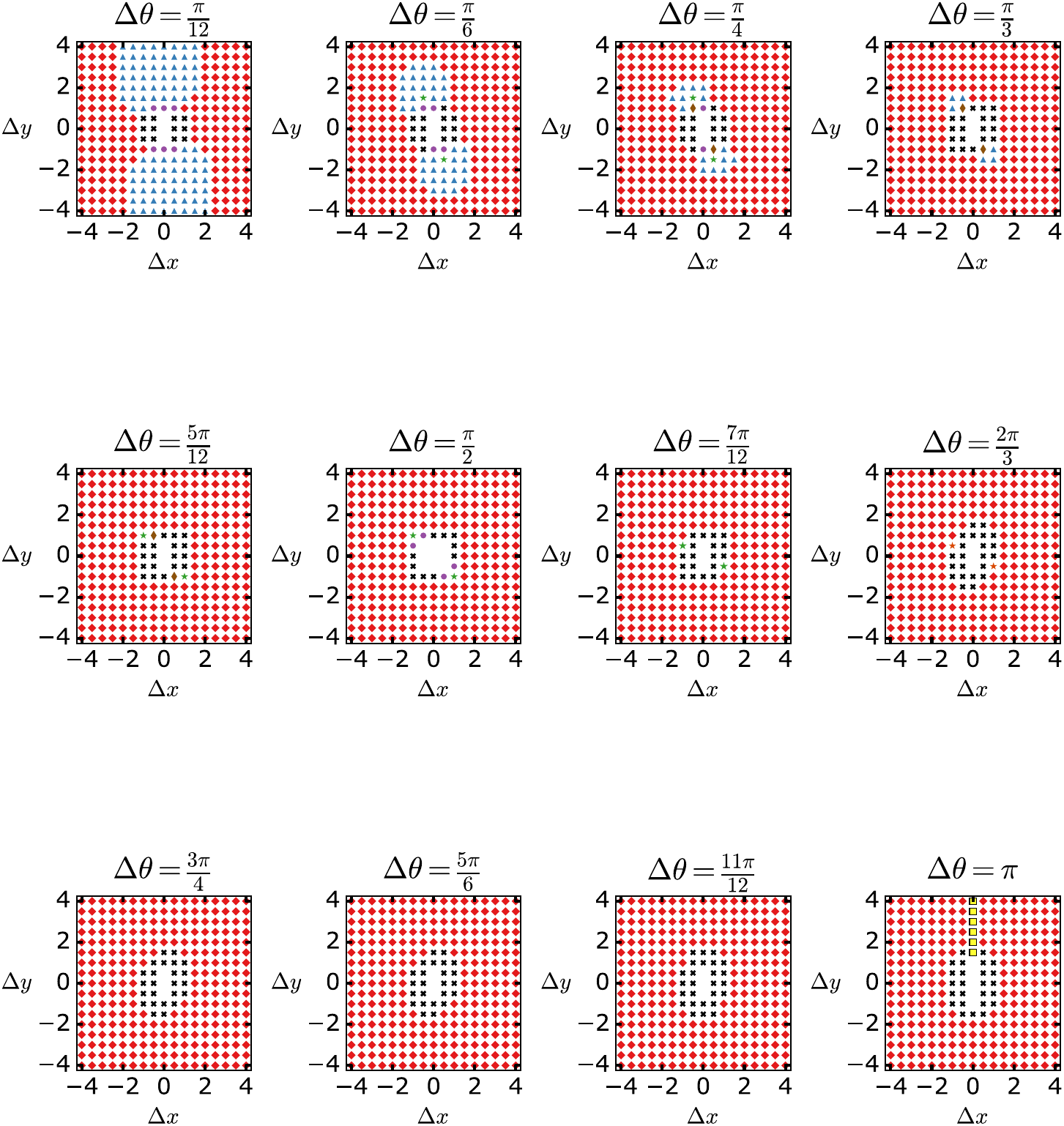}
    \caption{Phase diagrams for three-dimensional swimmers for $ \Delta \theta  >   0$.}
    \label{fig16:phaseDiagram3D_Full}
\end{figure}

The phase diagrams for the planar dynamics of a pair of three-dimensional source-sink swimmers with $\Delta \theta \neq 0$ are shown in Figure~\ref{fig16:phaseDiagram3D_Full} in increments of $\pi/12$. Symbols and colors are as in Figure~\ref{fig9:phaseDiagram3D}. Outside of the regions shown, the swimmers diverge from each other; that is, those regions of the phase diagrams are filled with red diamonds. The phase diagrams for $\Delta \theta < 0$ (not shown) are the same as for $\Delta \theta > 0$, but reflected across the $\Delta x = 0$ axis. 
\clearpage

\newpage
\section{\label{sec6:AppPD2D}Phase diagrams for two-dimensional model}

\begin{figure}[!h]
    \centering
    \includegraphics[scale=0.64]{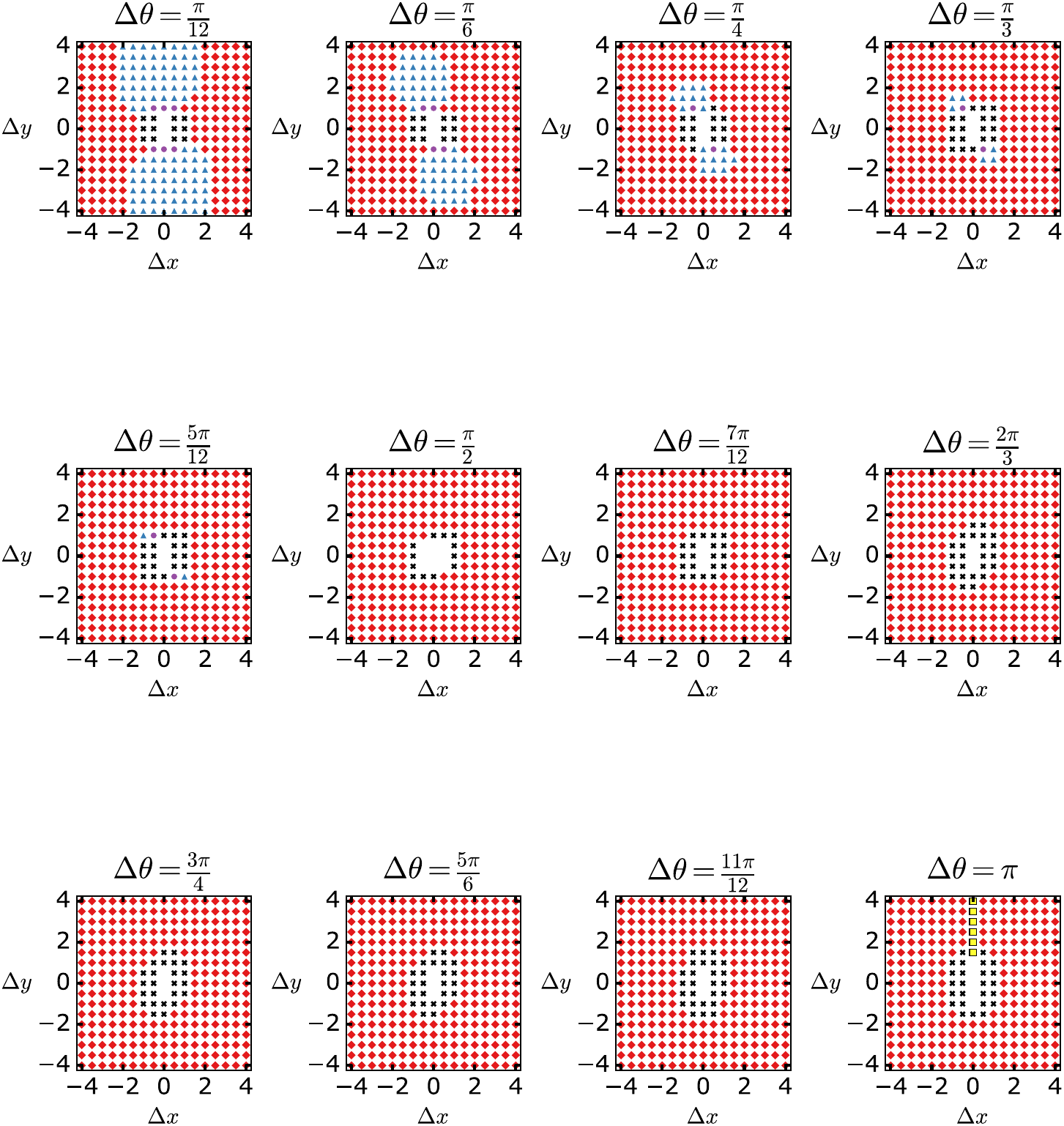}  
    \caption{Phase diagrams for two-dimensional swimmers for $ \Delta \theta  >   0$.}
    \label{fig17:phaseDiagram2D_Full}
\end{figure}

The phase diagrams for the planar dynamics of a pair of two-dimensional source-sink swimmers with $\Delta \theta \neq 0$ are shown in Figure~\ref{fig17:phaseDiagram2D_Full} in increments of $\pi/12$. Symbols and colors are as in Figure~\ref{fig13:phaseDiagram2D}. Outside of the regions shown, the swimmers diverge from each other; that is, those regions of the phase diagrams are filled with red diamonds. The phase diagrams for $\Delta \theta < 0$ (not shown) are the same as for $\Delta \theta > 0$, but reflected across the $\Delta x = 0$ axis. 

\end{document}